\documentclass{aastex631}
\usepackage{CJKutf8}
\usepackage[caption=false]{subfig}
\usepackage{graphicx}
\usepackage{amsmath}
\usepackage{extarrows}
\usepackage{multirow}
\usepackage{graphicx}
\usepackage{booktabs}
\usepackage{amsmath,amssymb}
\usepackage[T1]{fontenc}

\newcommand{\Ni}{{$^{56}$Ni }}
\newcommand{\Co}{{$^{56}$Co }}
\newcommand{\Msun}{M$_\odot$}

\begin{document}

\title{SN~2018gk Revisited: the Photosphere, the Central Engine, And the Putative Dust}

\correspondingauthor{Shan-Qin Wang \begin{CJK*}{UTF8}{gbsn}(王善钦)\end{CJK*}}
\email{shanqinwang@gxu.edu.cn}

\author{Tao Wang \begin{CJK*}{UTF8}{gbsn}(王涛)\end{CJK*}}
\affiliation{Guangxi Key Laboratory for Relativistic Astrophysics,
	School of Physical Science and Technology, Guangxi University, Nanning 530004,
	China}

\author{Shan-Qin Wang \begin{CJK*}{UTF8}{gbsn}(王善钦)\end{CJK*}}
\affiliation{Guangxi Key Laboratory for Relativistic Astrophysics,
School of Physical Science and Technology, Guangxi University, Nanning 530004,
China}

\author{Wen-Pei Gan \begin{CJK*}{UTF8}{gbsn}(甘文沛)\end{CJK*}}
\affiliation{Nanjing Institute of Astronomical Optics \& Technology, Nanjing 210042, China}

\author{Long Li \begin{CJK*}{UTF8}{gbsn}(李龙)\end{CJK*}}
\affiliation{Department of Astronomy, School of Physical Sciences, University of
Science and Technology of China, Hefei 230026, China}

\begin{abstract}

In this paper, we perform a comprehensive study for the physical properties of SN~2018gk which is a luminous type
IIb supernova (SN). We find that the early-time photospheric velocity vary from a larger value to a smaller value before
the photosphere reach a temperature floor. We generalize the photosphere modulus
and fit the multiband light curves (LCs) of SN~2018gk. We find that the \Ni mass model require
$\sim$ {0.90 \Msun} of \Ni which is larger than the derived ejecta mass ($\sim$ {0.10} \Msun).
Alternatively, we use the magnetar plus \Ni and the fallback plus \Ni models
to fit the LCs of SN~2018gk, finding that the two models can fit the
LCs. We favor the magnetar plus \Ni since the parameters are rather reasonable
({$M_{\rm ej} =$ 1.65 \Msun, $M_{\rm Ni}=0.05 $ \Msun} which is smaller
than the upper limit of the value of the \Ni mass can by synthesized by the
neutrino-powered core collapse SNe {$B=6.52\times10^{14}$} G which is comparable
to those of luminous and superluminous SNe studied in the literature, and $P_0=$ {10.42} ms
which is comparable to those of luminous SNe), {while the validity of the fallback plus \Ni
model depends on the accretion efficiency ($\eta$)}.
Therefore, we suggest that SN 2018gk might be a SN IIb mainly powered by a
central engine. Finally, we confirm the NIR excesses of the SEDs of SN~2018gk
at some epochs and constrain the physical properties of the putative dust using
the blackbody plus dust emission model.

\end{abstract}


\section{Introduction}
\label{sec:intro}

It is widely believed that massive stars whose zero-age main-sequence mass $M_{\rm ZAMS}$ are $\gtrsim 8.0 M_{\odot}$)
would explode as core-collapse supernovae (CCSNe; \citealt{Woosley2002}), leaving neutron star (NSs) or black hole (BHs).
Most CCSNe can be classified into types IIP, IIL, IIn, IIb, Ib, and Ic according to the spectra and light curves (LCs)
\citep{Filippenko1997}. In the past two decades, some new sub-classes of CCSNe, such as Ibn and Icn, are confirmed.
In general, {SNe} Ib, Ic, IIb, Ibn, and Icn can be collectively referred to as stripped-envelope SNe (SESNe;
\citealt{Clocchiatti1996}), since the hydrogen and/or helium envelopes of their progenitors were partly or completely stripped.

The early-time spectra of SNe IIb show hydrogen absorption lines, while their late-time
spectra do not show hydrogen lines, but show helium absorption lines \citep{Filippenko1988}.
These features make SNe IIb to be the transitional type
between {SNe II} whose spectra continue to show hydrogen absorption lines and SNe Ib whose spectra
have helium absorption lines. It is suggested that the progenitors stars of SNe IIb lost most of hydrogen envelopes through stellar wind
and/or binary interactions \citep{Podsiadlowski1992,Sravan2019,Yoon2010,Yoon2017}, leaving
$\sim$\,0.1\,M$\odot$ of hydrogen envelopes.

The peak absolute magnitudes ($M_{\rm peak}$) of most SNe IIb is dimmer than $-19.0$ mag.
However, there are few SNe IIb are overluminous, with $M_{\rm peak}$
brighter than $-19.0$ mag or even $-20.0$ mag which
are between those of superluminous supernovae (SLSNe, $M_{\rm peak}\lesssim -21$ mag, \citealt{Gal2012})
and normal SNe.\footnote{We note {that}, however, luminous SNe with $M_{\rm peak}$ being $-20$ to $-21$ mag
are also classified to be SLSNe in some literature.} For instance, the $M_{\rm peak}$ of
SNe IIb PTF10hgi, DES14X2fna, and SN~2018gk (ASASSN-18am) are
$\sim -20.42$ mag ($g-$ band, \citealt{Inserra2013})
\footnote{PTF10hgi had been classified to be a type Ic SLSN \cite{Inserra2013}, but \cite{Quimby2018}
suggest {that it is an} SLSN IIb, since its spectrum shows clear evidence of hydrogen and helium.},
$\sim -19.3$ mag ($g-$ band, \citealt{Grayling2021}),
and $\sim -19.7$ mag ($V-$ band, \citealt{Bose2021}, {hereafter B21}), respectively.

{This work focuses on} the physical properties of SN~2018gk.
SN~2018gk was discovered \citep{Brimacombe2018} by the All-Sky
Automated Survey for Supernovae (ASASSN; \citealt{Shappee2014,Kochanek2017}) on 2018 January
12.5 UT. The host galaxy of SN~2018gk is WISE J163554.27+400151.8 whose redshift $z$ is
$0.031010 \pm 0.000005$ \citep{SDSS2017}.
The spectra sequence of SN~2018gk indicate that it is
{an} SN IIb. B21 find that its $M_{\rm peak}$ in $V-$band is $\approx -19.7$ mag,
which is at least one {magnitude brighter} than those of most SNe IIb, and
comparable to those of luminous SNe IIb PTF10hgi and DES14X2fna.

B21 construct the bolometric LC and use the radiative diffusion
(\Ni {plus internal energy}) model
to fit it, finding that the required \Ni mass $M_{\rm Ni}$ is $\sim$ 0.4 \Msun.
They also use the magnetar plus \Ni model to fit the bolometric LC and find that
the initial period ($P_0$) and the magnetic field strength ($B$) of the magnetar are
$\sim$ 1.2 ms and $\sim 4 \times 10^{15}$ G, respectively.
As pointed {out} by B21, the value of $B$ is higher than those of
SLSNe I modeled by \cite{Nicholl2017} which are a few $10^{13}$ G to a few $10^{14}$ G.
B21 suggest that the extreme value of $B$, together with the
very short $P_0$ would extract rotational energy of up to $\sim 10^{52}$ erg
in the first 10 s after the explosion, thus significantly affecting SN shock dynamics.
{Furthermore, B21 point out that t}he magnetar do not significantly
reduce $M_{\rm Ni}$ since the \Ni mass derived
by the magnetar plus \Ni model is $\sim$ 0.3 \Msun.
The \Ni masses derived by the two models are larger than the upper limit
($\sim$ 0.2 \Msun) of the \Ni mass that can be synthesized by neutrino-power
mechanism \citep{Sukhbold2016,Ertl2020}.
{\cite{Soker2021} fit the $V-$ band data of SN~2018gk using a two-components
bipolar model. The photometry in all other bands have not been used.}

Therefore, we suggest that the energy source of SN~2018gk is still elusive
{and deserve further study. In this paper, we research the energy source model,
the physical parameters of the models as well as the properties of the possible dust of SN~2018gk.}
In Section \ref{sec:modeling}, we model the multi-band LCs for SN~2018gk using three models
(the \Ni model, the magnetar plus \Ni model, and the fallback plus \Ni model).
In Section \ref{sec:dust}, we study the NIR excesses of SN~2018gk and constrain the
physical properties of the putative dust.
In Section \ref{sec:discussion}, we discuss our results. We draw some
conclusions in Section \ref{sec:conclusions}.
Throughout the paper, we assume $\Omega_m = 0.315$, $\Omega_\Lambda = 0.685$,
and H$_0 = 67.3$\,km\,s$^{-1}$\,Mpc$^{-1}$ \citep{Planck2014}. The values of the Milky Way
reddening ($E_\mathrm{B-V}$) from \cite{Schlafly2011}.

\section{Modeling the Multi-band Light Curves of SN~2018gk}
\label{sec:modeling}

B21 construct and fit the bolometric LC of SN~2018gk.
It should be noted that, however, the upper limits at two very early epochs in $V-$ band
and six data in $V-$ and $g-$ bands at four early epochs cannot be included in
the constructed LC. Therefore, modeling the multi-band LCs might get more reliable
results and pose more stringent constraints on the physical properties of SN~2018gk.

The bolometric luminosity of SNe powered is (see e.g., \citealt{Arnett1982,Chatzopoulos2012,Wang2015a,Wang+Gan2022})
\begin{eqnarray}
L_{\rm SN}(t)=\frac{2}{\tau_{m}}e^{-\frac{t^{2}}{\tau_{m}^{2}}}\int_0^t e^{\frac{t'^{2}}{\tau_{m}^{2}}}
\frac{t'}{\tau_{m}}L_{\rm input}(t')\left(1-e^{-\tau_{\gamma\,(\rm e^+)}(t)}\right)dt',
\label{equ:lum}
\end{eqnarray}
here, $\tau_{m}=({2\kappa M_{\rm ej}}/{\beta_{\rm SN} v_{\rm sc}c})^{1/2}$ is the diffusion timescale {\citep{Arnett1982}},
$\kappa$ is the optical opacity of the ejecta which is usually assumed to be $0.05-0.20$ cm$^2$ g$^{-1}$,
{$M_{\rm ej}$ is the SN ejecta mass, $v_{\rm sc}$ is the scale velocity of the SN \citep{Arnett1982}},
$c$ is the speed of light, $\beta_{\rm SN} \simeq 13.8$ is a constant \citep{Arnett1982};
$\tau_{\gamma\,(\rm e^+)}(t)=t_{0,\gamma\,(\rm e^+)}^2/ t^2$ is the optical depth to
$\gamma$-rays \citep{Chatzopoulos2009,Chatzopoulos2012} or {positrons,
$\kappa_{\gamma,\rm Ni}$ and $\kappa_{\rm e^+}$ are respectively the opacity of the $\gamma$-rays from \Ni cascade decay
and the positron from \Co decay, $t_{0,\gamma\,(\rm e^+)}=({3\kappa_{\gamma\,(\rm e^+)}M_{\rm ej}}/{4\pi v^2})^{1/2}$
is $\gamma$-ray (positron) trapping parameter;}
$L_{\rm input}(t)$ is the power function of the energy source(s).

As multi-band LC fitting is used, a model must be included for the spectral energy distributions (SEDs).
The SEDs of the SNe can be described by the UV absorbed blackbody model \citep{Nicholl2017,Prajs2017},
\begin{eqnarray}
	F_{\nu} =
	\left\{
	\begin{array}{lr}
		\big(\frac{\lambda}{\lambda_{\rm CF}}\big)^{\beta'}(2 {\pi} h{\nu}^3/c^2)(e^{\frac{h{\nu}}{k_{\rm b}T_{\rm ph}}}-1)^{-1}\frac{R_{\rm ph}^2}{D_L^2},
		\ & \quad \lambda \leq \lambda_{\rm CF} \\
		(2 {\pi} h{\nu}^3/c^2)(e^{\frac{h{\nu}}{k_{\rm b}T_{\rm ph}}}-1)^{-1}\frac{R_{\rm ph}^2}{D_L^2},
		\ & \quad \lambda > \lambda_{\rm CF} \\
	\end{array}
	\right.
	\label{equ:SED}
\end{eqnarray}
$T_{\rm ph}$ is the temperature of the SN photosphere, $R_{\rm ph}$ is the
radius of the SN photosphere,
$D_{\rm L}$ is the luminosity distance of the SN,
$\lambda_{\rm CF}$ is the cutoff wavelength,
$\beta'$ is a dimensionless free parameter \citep{Yan2020}.

\begin{figure}[tbph]
\begin{center}
\includegraphics[width=0.48\textwidth,angle=0]{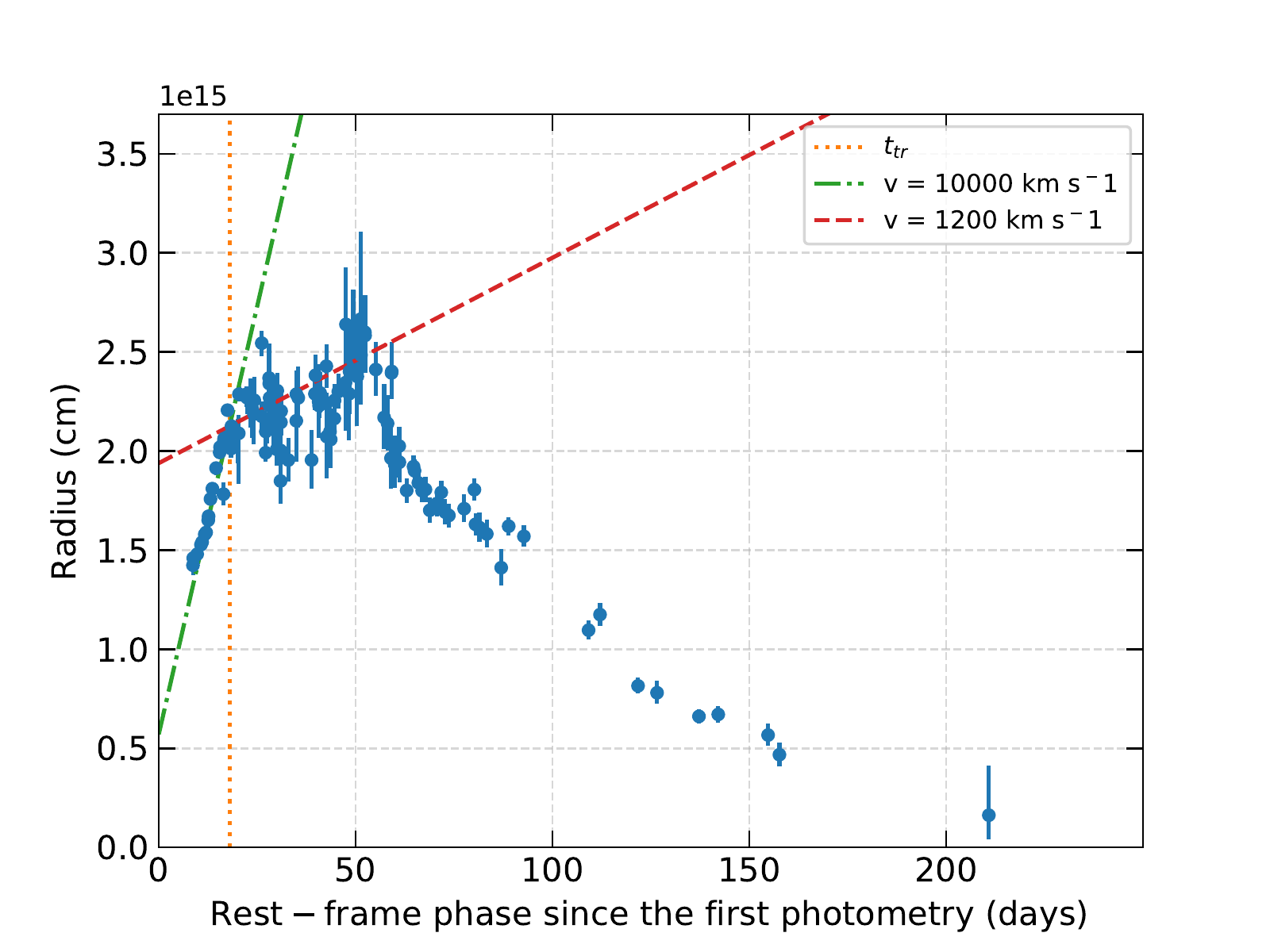}
\includegraphics[width=0.48\textwidth,angle=0]{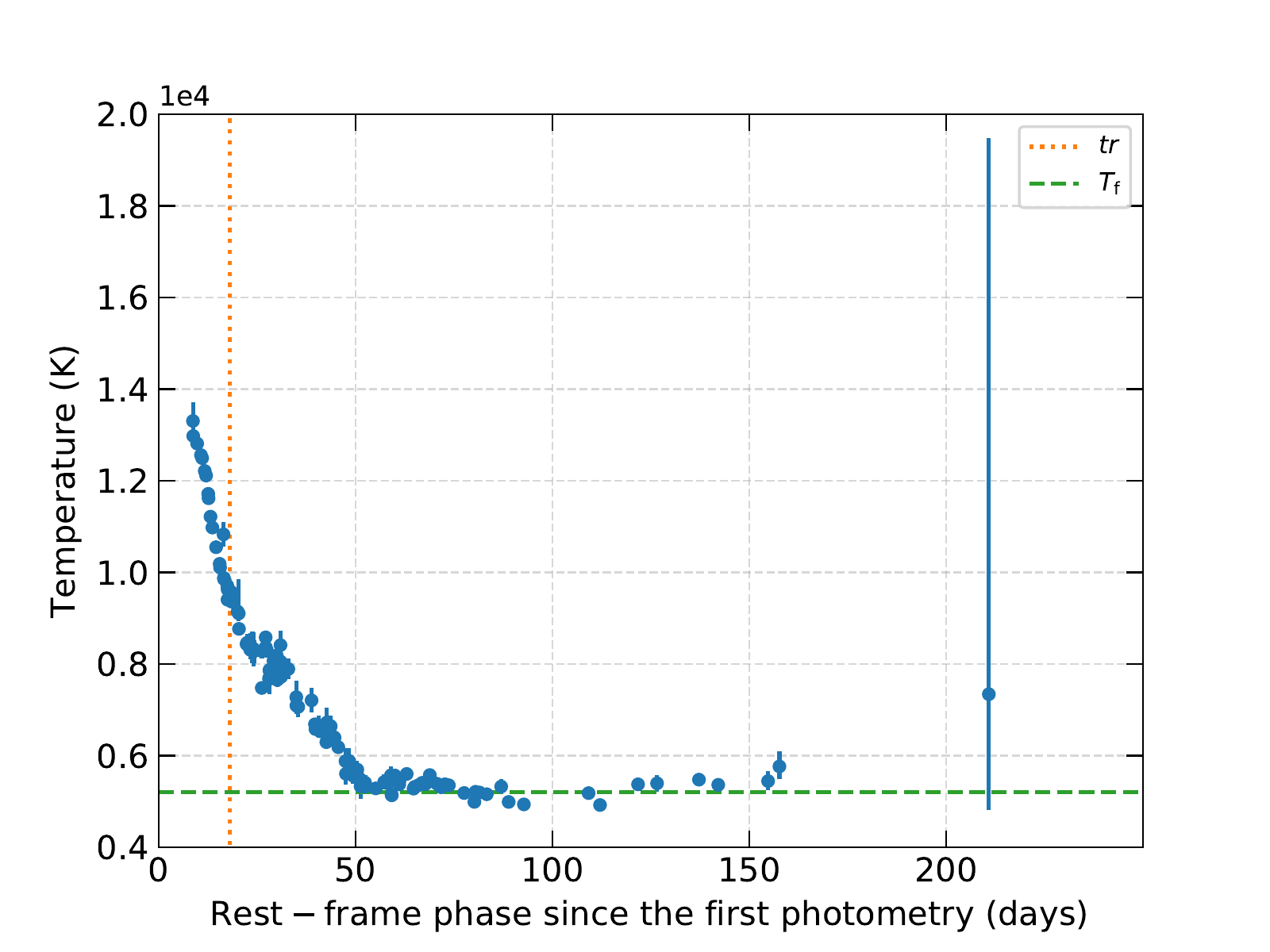}
\end{center}
\caption{The radius evolution and the temperature evolution derived by the fits for the SEDs at different epochs.}
\label{fig:RandT_point}
\end{figure}

In general, the photosphere evolution can be described by \cite{Nicholl2017} which
assumes that the early photosphere expansion velocity ($v_{\rm ph}$)
of SNe is constant and the late-time temperature is a constant ($T_{\rm f}$).
We find, however, that the curve of early photosphere expansion velocity
can be divided into two episodes having different velocities ($v_{\rm ph1}$
and $v_{\rm ph2}$, see {the green dash-dotted line and the red dashed line} of
Figure \ref{fig:RandT_point}). The {transition time} when $v_{\rm ph1}$
became $v_{\rm ph2}$ is denoted as $t_{\rm tr}$ {which is respective to the time of the first
photometry (see {the orange vertical dotted line} of Figure \ref{fig:RandT_point}).
The time interval between the explosion date and the the time when $v_{\rm ph1}$
became $v_{\rm ph2}$ ($t_{\rm 1}$) is $t_{\rm tr}-t_{\rm shift}$; here, $t_{\rm shift}$
is the explosion time relative to the first photometry.} Therefore, we generalize the photosphere modulus as follow,
\begin{eqnarray}
	R_{\rm ph}(t) =
	\left\{
	\begin{array}{lr}
		v_{\rm ph1} t, &
		\qquad t\le t_1 \quad \mathrm{and} \quad \left(\frac{L_{\rm SN}(t)}{4 \pi \sigma (v_{\rm ph1}t)^2}\right)^{\frac{1}{4}} > T_{\rm f}\\
		v_{\rm ph1} t_1 + v_{\rm ph2} (t-t_1), &
		\qquad t > t_1 \quad \mathrm{and} \quad \left(\frac{L_{\rm SN}(t)}{4 \pi \sigma (v_{\rm ph1}t_1 + v_{\rm ph2}t)^2}\right)^{\frac{1}{4}} > T_{\rm f}\\
		\left(\frac{L(t)}{4 \pi \sigma T_{\rm f}^4}\right)^{\frac{1}{2}},&
		\left(\frac{L(t)}{4 \pi \sigma (v_{\rm ph1}t_1 + v_{\rm ph2}t)^2}\right)^{\frac{1}{4}} \le T_{\rm f} \\
	\end{array}
	\right.
	\label{equ:2-v-R}
\\
	T_{\rm ph}(t) =
	\left\{
	\begin{array}{lr}
		\left(\frac{L_{\rm SN}(t)}{4 \pi \sigma v_{\rm ph1}^2 t^2}\right)^{\frac{1}{4}},\ &
		\qquad t\le t_1 \quad \mathrm{and} \quad \left(\frac{L_{\rm SN}(t)}{4 \pi \sigma (v_{\rm ph1}t)^2}\right)^{\frac{1}{4}} > T_{\rm f}\\
		\left(\frac{L_{\rm SN}(t)}{4 \pi \sigma (v_{\rm ph1}t_1 + v_{\rm ph2}t)^2}\right)^{\frac{1}{4}},\ &
		\qquad t > t_1 \quad \mathrm{and} \quad \left(\frac{L_{\rm SN}(t)}{4 \pi \sigma (v_{\rm ph1}t_1 + v_{\rm ph2}t)^2}\right)^{\frac{1}{4}} > T_{\rm f}\\
		T_{\rm f},&
		\left(\frac{L(t)}{4 \pi \sigma (v_{\rm ph1}t_1 + v_{\rm ph2}t)^2}\right)^{\frac{1}{4}} \le T_{\rm f} \\
	\end{array}
	\right.
	\label{equ:2-v-T}
\end{eqnarray}

\subsection{Modeling the multi-band LCs of SN~2018gk using the \Ni model}
\label{subsec:ni}

We first fit the multi-band LCs of SN~2018gk using the the \Ni model. {Throughout this paper,
$v_{\rm sc}$ is assumed to be $v_{\rm ph1}$. To include the energy deposited by
$\gamma$-rays and positrons, we adopt a more complicated equation below (see, e.g., \citealt{Valenti2008})}
\begin{eqnarray}
	L_{\rm input}(t)\left(1-e^{-\tau_{\gamma\,(\rm e^+)}(t)}\right) = S^{\rm Ni}(\gamma) + S^{\rm Co}(\gamma)+ S^{\rm Co}_{\rm e^+}(\gamma) +S^{\rm Co}_{\rm e^+}(\rm KE)
\label{equ:input-Ni-positron}
\end{eqnarray}
{here, $S^{\rm Ni}(\gamma) = \epsilon_{\rm Ni}M_{\rm Ni}e^{-{t}/{\tau_{\rm Ni}}}\left(1-e^{-\tau_{\gamma}(t)}\right)$
is the source energy of the \Ni decay, $M_{\rm Ni}$ is the initial mass of \Ni;
$\epsilon_{\rm Ni}=3.9\times 10^{10}$~erg~s$^{-1}$~g$^{-1}$ \citep{Cappellaro1997,Sutherland1984}
is the energy generation rate of \Ni, $\tau_{\rm Ni}=$ 8.8 days is the lifetime of the \Ni;
$S^{\rm Co}(\gamma) = 0.81 \mathcal{E}_{\rm Co} (1-e^{-({{t_{0,\gamma}}/{t}})^2})$,
$S^{\rm Co}_{\rm e^+}(\gamma) = 0.164 \mathcal{E}_{\rm Co} (1-e^{-({{t_{0,\gamma}}/{t}})^2})(1-e^{-({{t_{0,\rm e^{+}}}/{t}})^2})$, and
$S^{\rm Co}_{\rm e^+}(\rm KE) = 0.036 \mathcal{E}_{\rm Co} (1-e^{-({{t_{0,\rm e^{+}}}/{t}})^2})$
are respectively the source energy of the \Co decay, the energy deposited by
the $\gamma$-rays produced in the positron annihilation, and the source energy due to the kinetic energy of the
positrons \citep{Valenti2008}, where
$\mathcal{E}_{\rm Co} = \epsilon_{\rm Co}M_{\rm Ni}e^{-{t}/{\tau_{\rm Co}}}-e^{-{t}/{\tau_{\rm Ni}}}$
is the rate of energy production by the \Co decay, $\epsilon_{\rm Co}=6.8 \times 10^{9}$ erg~s$^{-1}$~g$^{-1}$ \citep{Maeda2003}
is the energy generation rate of \Co, $\tau_{\rm Co}=$111.3 days is the lifetime of \Co}.

By combining Equations \ref{equ:lum}, \ref{equ:SED}, \ref{equ:2-v-R}, \ref{equ:2-v-T}, and \ref{equ:input-Ni-positron},
we can construct the \Ni model used to fit the multi-band LCs of SN~2018gk.
Considering the fact that the rise time is rather short, we fixed the
value of $\kappa$ to be 0.05 cm$^2$g$^{-1}$ throughout this paper to obtain
a relative larger ejecta mass.
The definitions, the units, and the priors of the parameters of the \Ni model are listed in Table \ref{table:parameters-ni}.
{We note that $\kappa_{\gamma,\rm Ni}$ and $\kappa_{\rm e^{+}}$ are also free parameters in our model}.\footnote{{It it believed
	that the value of $\kappa_{\gamma,\rm Ni}$ is determined by the $\gamma$-ray energy $E_{\gamma}$
(0.314 and 0.814 Mev for \Ni decay to \Co and then to $^{56}$Fe) and the number of baryons per electron $Y_{\rm e}$
($\sim$ 0.5). \cite{Swartz1995} find MCMC that $\kappa_{\gamma,\rm Ni}$ is about $0.06^{+0.01}_{-0.01}Y_{\rm e}$ at $E_{\gamma} > 0.1$ MeV.
Therefore, $\kappa_{\gamma,\rm Ni}$ is usually set to be 0.027 cm$^2$g$^{-1}$ (e.g., \citealt{Cappellaro1997,Maeda2003,Mazzali2000}).
However, there is evidence that $\kappa_{\gamma,\rm Ni}$ of some SNe has larger values. For instance,
based on the the parameters derived by \cite{Nagy2016} for SN~1987A ($A_{\rm g}=2.7\times10^{5}$ day$^2$,
$M_{\rm ej} =$ 8.6 \Msun, $E_{\rm k} = 1.17 \times 10^{51}$ erg), we find that $\kappa_{\gamma,\rm Ni} \sim $ 0.11 cm$^2$g$^{-1}$.
So we assume $\kappa_{\gamma,\rm Ni}$ is a free parameter whose lower limit is 0.027 cm$^2$g$^{-1}$.
The values of $\kappa_{\rm e^{+}}$ are usually set to be 7 cm$^2$g$^{-1}$ or 10 cm$^2$g$^{-1}$ \citep{Milne1999}.
\cite{Cappellaro1997} argued that 7 cm$^2$g$^{-1}$ is the best value of $\kappa_{\rm e^{+}}$;
however, they derived this value based on the assumption that $\kappa_{\gamma,\rm Ni}$ is 0.027 cm$^2$g$^{-1}$.
We do not adopt this value, since $\kappa_{\gamma,\rm Ni}$ is set to be a free parameter in this
work. In fact, the value of $\kappa_{\rm e^{+}}$ depends on the value of the number of free electrons per nucleus $\chi_e$
(the ionization fraction, see, e.g., \citealt{Milne1999}): for $\chi_e$ = 0.01, $\kappa_{\rm e^{+}}$ = 4 cm$^2$g$^{-1}$;
for $\chi_e$ = 3, $\kappa_{\rm e^{+}}$ = 14 cm$^2$g$^{-1}$.
We assume that $\kappa_{\rm e^{+}}$ is also a free parameter whose lower limit and upper limit
are 4 cm$^2$g$^{-1}$ and 14 cm$^2$g$^{-1}$, respectively.}}

The Markov Chain Monte Carlo (MCMC) using the \texttt{emcee} Python package \citep{Foreman-Mackey2013}
is adopted to get the best-fitting parameters and the
1-$\sigma$ range of the parameters. We employ 20 walkers, each walker runs 30,000 steps
for the \Ni model. The 1-$\sigma$ uncertainties are corresponding to the 16th and
84th percentiles of the posterior samples.

\begin{figure}[tbph]
\begin{center}
\includegraphics[width=0.7\textwidth,angle=0]{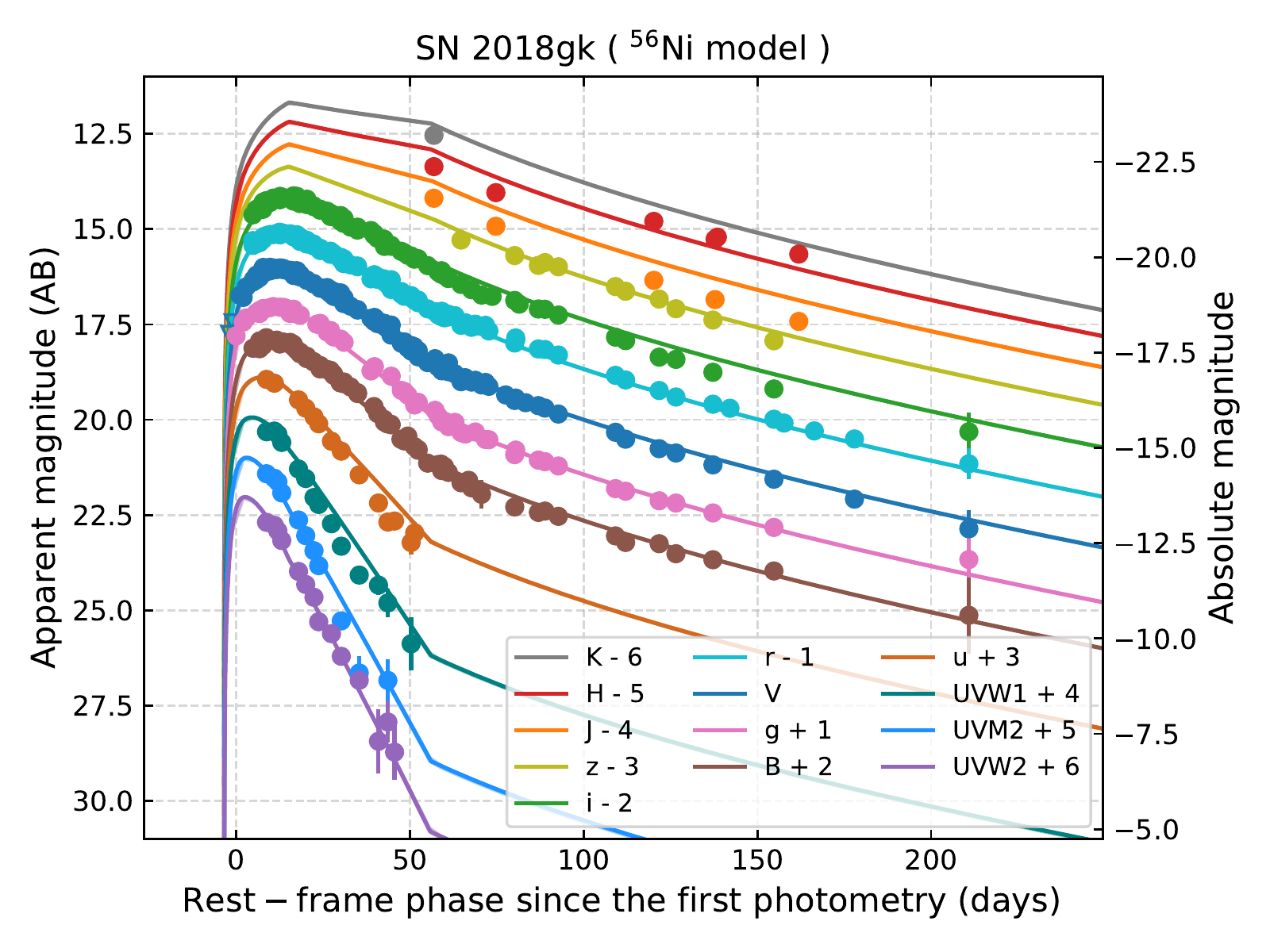}
\end{center}
\caption{The best fits (the solid curves) of the multi-band LCs
	of SN~2018gk using the \Ni model. The shaded regions indicate 1-$\sigma$ bounds
	of the parameters. The data are from Table 1 of B21, triangles represent upper limits.}
\label{fig:multibandfits-ni}
\end{figure}

The fit of the \Ni model is shown in Figure \ref{fig:multibandfits-ni}. The parameters and the corresponding
corner plot are presented in Table \ref{table:parameters-ni} and Figure \ref{fig:corner_Ni}, respectively.
{The relevant parameters of the \Ni model are $M_{\rm ej} =$ 0.10 \Msun,
$t_{\rm tr}$ $=$ 14.02 days, $v_{\rm ph1}=1.52 \times 10^9$ cm s$^{-1}$, $v_{\rm ph2}=0.16 \times 10^9$ cm s$^{-1}$,
$M_{\rm Ni} =$ 1.59 \Msun, $T_{\rm f}=4921.90 $ K, {$t_{\rm shift} = -$1.76 days
(the offset of the best-fitting explosion epoch with respect to the explosion epoch adopted by B21 is 1.36 days), and} $\kappa_{\gamma,\rm Ni} =$ 0.35 cm$^2$g$^{-1}$.
Using these parameters, we find that the value of $\gamma$-ray trapping parameter $t_{0,\gamma,\rm Ni}$ is 31.25 days.}
We find that the derived \Ni mass (1.59 \Msun) is significantly larger than the derived ejecta mass (0.10 \Msun).

It should be noted that, however, \cite{Arnett1982}'s model we adopt above might overestimate the
\Ni masses of SESNe (including IIb). We therefore calculate the more accurate value of \Ni mass of SN~2018gk
using the Equation derived by \cite{Khatami2019}
\begin{eqnarray}
	M_{\rm Ni} = \frac{L_{\rm peak} \beta^2 t_{\rm peak}^2 }{2 \epsilon_{\rm Ni} \tau_{\rm Ni}^2} \left( \left(1 - \frac{\epsilon_{\rm Co}}{\epsilon_{\rm Ni}} \right)\times (1 -
	(1 + \beta t_{\rm peak}/\tau_{\rm Ni})e^{-\beta t_{\rm peak}/\tau_{\rm Ni}}) + \frac{\epsilon_{\rm Co} \tau_{\rm Co}^2}{\epsilon_{\rm Ni} \tau_{\rm Ni}^2} \left(1 - (1+\beta t_{\rm peak}/\tau_{\rm Co} )e^{-\beta t_{\rm peak}/\tau_{\rm Co}} \right)  \right)^{-1}
	\label{equ:newNimass}
\end{eqnarray}
The value of $\beta$ is $\sim$0.78 for SNe IIb \citep{Afsariardchi2021}.

Using the peak luminosity ($L_{\rm peak}$) and the rise time ($t_{\rm peak}$)
of the bolometric LC produced by the best-fitting parameters of the multi-band
LC fit, we find that the $M_{\rm Ni}$ value derived by Equation \ref{equ:newNimass}
is 0.90 \Msun. While this value is lower than the value derived by \cite{Arnett1982}'s model,
it is significantly
{higher than the value of \Ni derived by B21 using the \cite{Katz2013} integral method} and
and is still larger than the mass of the eject.
This indicate that the \Ni model cannot account for the LCs of SN~2018gk.

\begin{figure}[tbph]
	\begin{center}
		\includegraphics[width=0.7\textwidth,angle=0]{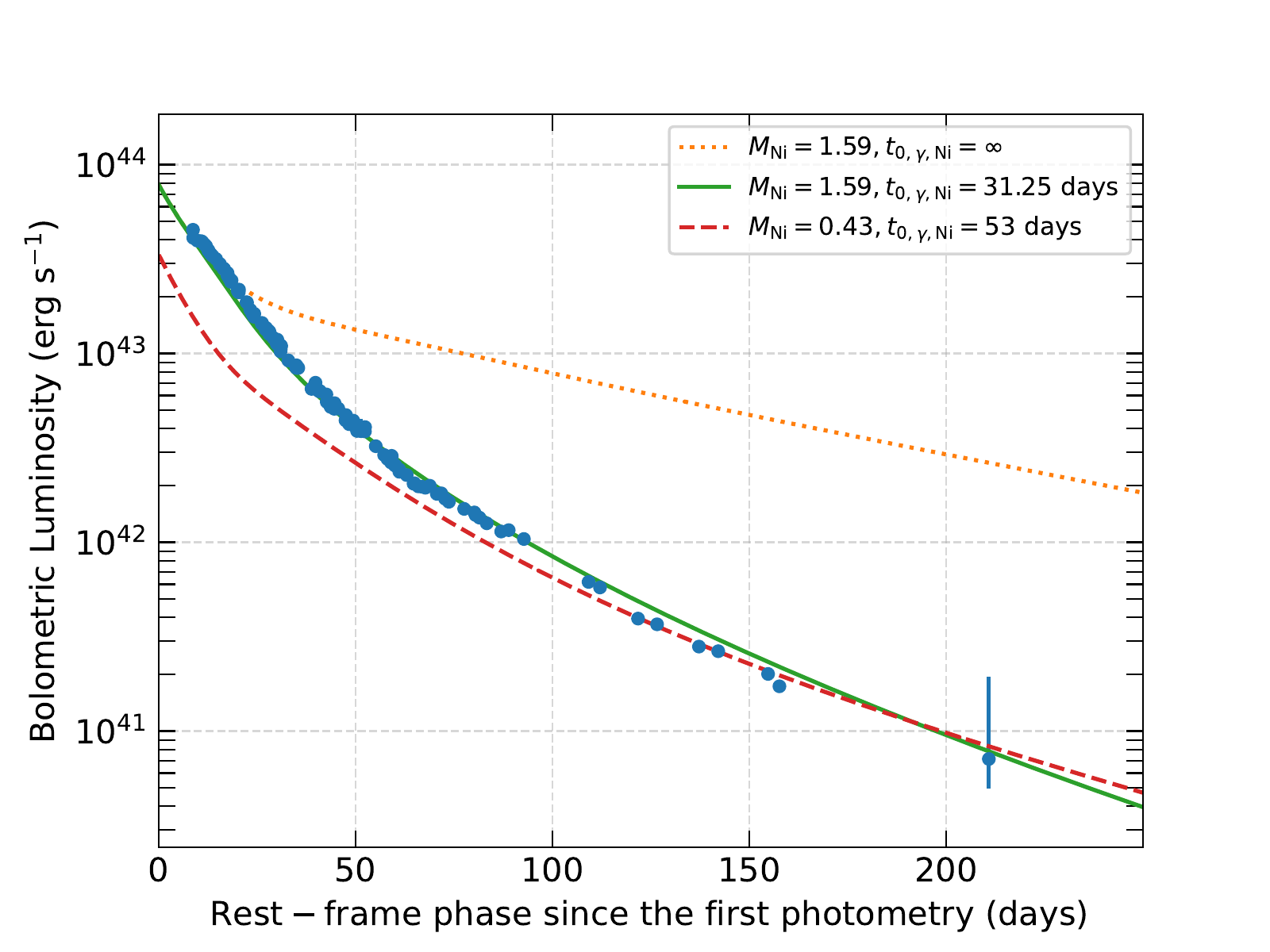}
	\end{center}
	\caption{{Our synthesized bolometric LC of SN~2018gk (blue points), the power injection function curve
derived from the parameters of B21 ($M_{\rm Ni} =$ 0.43\Msun, $t_{0,\gamma,\rm Ni}$ = 53 days, {$t_{\rm shift} = -$0.4 days, the} red dashed line),
as well as the power injection function curve derived from our parameters
($M_{\rm Ni} =$1.59 \Msun, $t_{0,\gamma,\rm Ni}$ = 31.25 days, {$t_{\rm shift} = -$1.76 days, the} green solid line).
For comparison, the curve derived from $M_{\rm Ni} =$ 1.59 \Msun and a infinite $t_{0,\gamma,\rm Ni}$ is also
plotted ({the} orange dotted line).}}
	\label{fig:late-time-fit}
\end{figure}

{In Figure \ref{fig:late-time-fit}, we plot our synthesized bolometric LC of
SN~2018gk, the power injection function curve
derived from the parameters of B21 ($M_{\rm Ni} =$ 0.43\Msun, $t_{0,\gamma,\rm Ni}$ = 53 days),
as well as the power injection function curve
derived from our parameters ($M_{\rm Ni} =$1.59 \Msun, $t_{0,\gamma,\rm Ni}$ = 31.25 days).
For comparison, the curve derived from $M_{\rm Ni} =$ 1.85 \Msun and a infinite $t_{0,\gamma,\rm Ni}$ is also
plotted.}

{We find that both the curves derived by B21's parameters and our parameters
can fit the late-time (>75 days) can match the synthesized bolometric LC, while
the curve with infinite $t_{0,\gamma,\rm Ni}$ cannot fit the bolometric LC.
It should be noted that, however, the curve using the B21's parameters
is much lower than the bolometric luminosity of the SN~2018gk around the peak,
indicating that, in B21's two-component diffusion model,
the early-time LC is (mainly) powered by the internal energy.}

{In fact, the \Ni masses derived from the peak luminosity of SLSNe and luminous
SNe are usually significantly higher than those derived from the tails (except
for the cases in which the SLSNe is pair instability SNe candidates), see, e.g.,
Table 3 of \cite{De2018}.
The \Ni masses derived from tails are most reliable. Therefore, the peak luminosity
of SLSNe and luminous SNe are usually assumed to be mainly powered by the energy
sources (magnetar, ejecta-CSM interaction, fallback, or internal energy)
rather than \Ni heating.}
	
{SN~2018gk is a luminous SN having observational properties similar to those of
SLSNe and luminous SNe. The large discrepancy between the inferred \Ni masses
derived from the peak (see above) and the tail (inferred by B21 using \citealt{Katz2013})
is also indicative of the necessity of introducing another energy source to account for
the luminous peak.}

\subsection{Modeling the multi-band LCs of SN~2018gk using the magnetar Plus \Ni model}
\label{subsec:magmodel+ni}

{Although B21 show that the two-component diffusion model taking
the internal energy and \Ni heating can fit the bolometric LC of SN~2018gk, the derived
\Ni mass ($\sim$ 0.43\Msun) is higher than the upper limit
($\sim$ 0.2 \Msun) of the \Ni mass that can be synthesized by neutrino-power
mechanism.
To reduce the required \Ni, B21 use the magnetar plus \Ni model to fit the
bolometric LC of SN~2018gk. They find that the model \Ni only slightly reduce the
\Ni mass to 0.3 \Msun which is still larger than 0.2 \Msun. Moreover, they find that
the derived parameters of the magnetars are very extreme: the initial period ($P_0$)
and the magnetic field strength ($B$) of the magnetar are $\sim$ 1.2 ms and
$\sim 4 \times 10^{15}$ G, respectively.
These values are inconsistent with the typical values of the
magnetars supposed to power the
LCs of luminous SNe (see, e.g., \citealt{Wang2015b,Gomez2022})
and SLSNe (see, e.g., \citealt{Nicholl2017,Liu2017}).}

{Hence, we check the magnetar plus \Ni model by fitting the LCs of SN~2018gk.}
The power function $L_{\rm input}(t)$ of the magnetar we adopt is
\begin{equation}
	L_{\rm input}(t)=L_{\rm input,mag}(t) = \frac{E_{p}}{\tau_{p}} \frac{1}{(1+t/\tau_{p})^{2}},
	\label{equ:input-mag}
\end{equation}

where $E_{p}\simeq({1}/{2})I_{\rm mag}\Omega_{\rm mag}^2$
is the rotational energy of a magnetar,
{$\tau_{p} = {6I_{\rm mag} c^3}/{B^2R_{\rm mag}^6\Omega_{\rm mag}^2}$}
is the spin-down timescale of the magnetar \citep{Kasen2010}.
$I_{\rm mag} = (2/5)M_{\rm mag}R_{\rm mag}^2$ is the moment of inertia
of a magnetar, {$\Omega_{\rm mag} = 2 \pi / P_0$
is initial angular velocity ($P_0$ is initial period of the magnetar),
$B$ is the magnetic field strength of the magnetar.
The mass ($M_{\rm mag}$) and the radius ($R_{\rm mag}$) of
the magnetar are usually set to be 1.4 \Msun and 10 km, respectively;
using the two values, the} canonical value of $I_{\rm mag}$ is
$\sim$ $10^{45}$~g~cm$^{2}$ (see, e.g., \citealt{Woosley2010}).
The $\gamma$-ray opacity $\kappa_{\gamma}$ in Equation \ref{equ:lum}
become $\kappa_{\gamma,\rm mag}$.

By combining Equations \ref{equ:lum}, \ref{equ:SED}, \ref{equ:2-v-R}, \ref{equ:2-v-T}, \ref{equ:input-Ni-positron}, and \ref{equ:input-mag},
the magnetar plus \Ni model can be constructed.
The definitions, the units, and the priors of the parameters of the magnetar plus \Ni model are listed in Table \ref{table:parameters-mag}.
The MCMC using the \texttt{emcee} Python package is adopted to get the best-fitting parameters and the
1-$\sigma$ range of the parameters. We employ 20 walkers, each walker runs 30,000 steps
for the magnetar plus \Ni model. The 1-$\sigma$ uncertainties are corresponding to the 16th and
84th percentiles of the posterior samples.

\begin{figure}[tbph]
\begin{center}
\includegraphics[width=0.7\textwidth,angle=0]{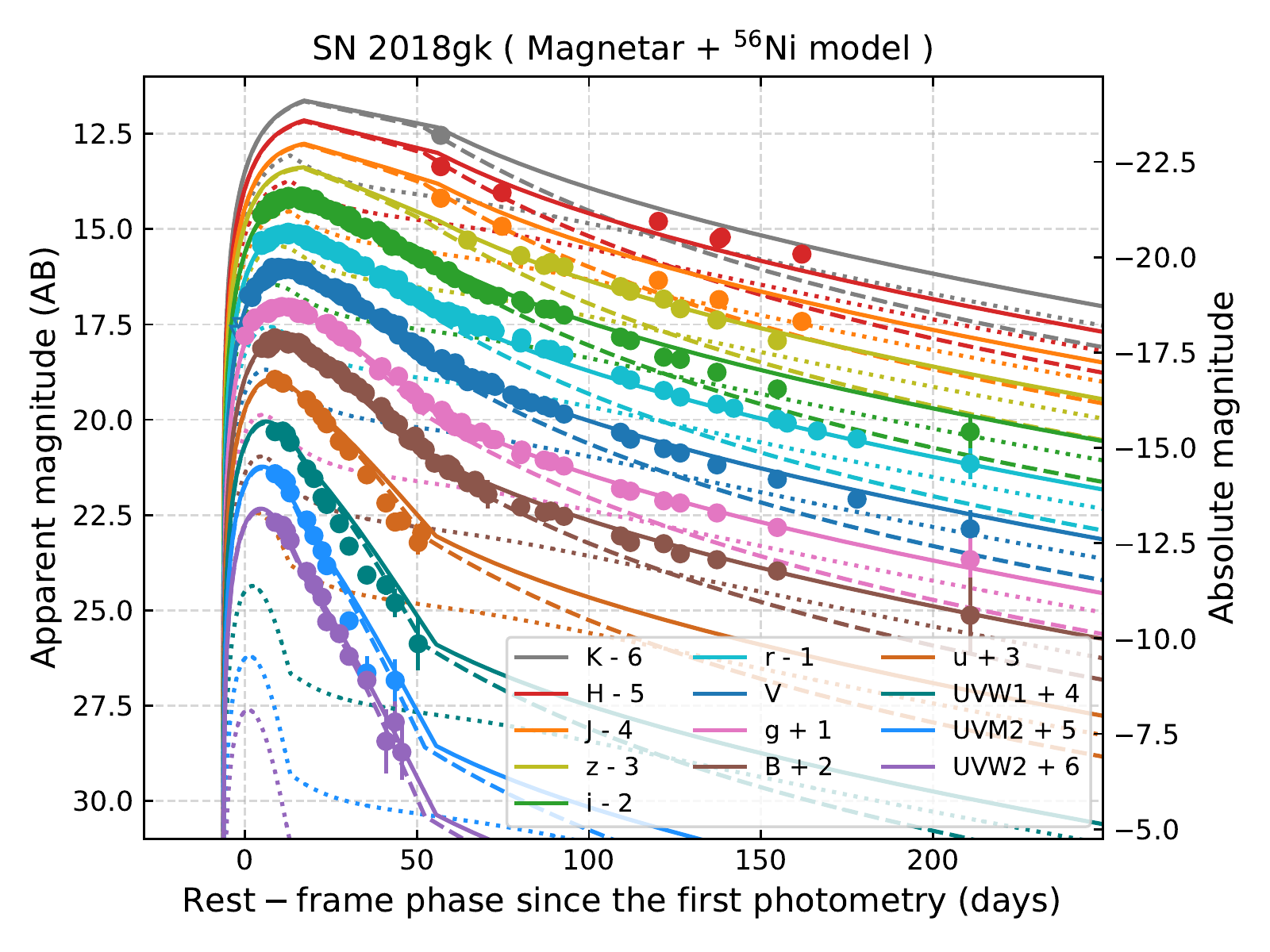}
\end{center}
\caption{The best fits (the solid curves) of the multi-band LCs
	of SN~2018gk using the magnetar plus \Ni. The shaded regions indicate 1-$\sigma$ bounds
	of the parameters. The dotted lines and dashed lines
	are the LCs powered by the \Ni and the magnetar, respectively.
	The data are from Table 1 of B21, triangles represent upper limits.}
\label{fig:multibandfits-mag}
\end{figure}

The fit of the magnetar plus \Ni model is shown in Figure \ref{fig:multibandfits-mag}.
The parameters {and the corresponding corner plot are} presented in
Table \ref{table:parameters-mag} {and Figure \ref{fig:corner_Mag+Ni}, respectively}.
The relevant parameters of the magnetar plus \Ni model are
$M_{\rm ej} = 1.65$ \Msun, $P_0= 10.42$ ms, $B= 6.52\times10^{14}$ G, $M_{\rm Ni} = 0.08$ \Msun,
$t_{\rm tr} = 17.15$ days, $v_{\rm ph1}=1.07 \times 10^9$ cm s$^{-1}$, $v_{\rm ph2}=0.11 \times 10^9$ cm s$^{-1}$,
$T_{\rm f}=5254.04 $ K, {$t_{\rm shift} = -$6.35 days (the offset of the best-fitting explosion epoch
with respect to the explosion epoch adopted by B21 is 5.95 days),
$\kappa_{\gamma,\rm Ni} =$ 0.19 cm$^2$g$^{-1}$ ($t_{0,\gamma,\rm Ni}$ = 132.17 days),
and $\kappa_{\gamma,\rm mag} =$ 0.028 cm$^2$g$^{-1}$ ($t_{0,\gamma,\rm mag}$ = 50.83 days).}

{We find that}, using the same scenario (the magnetar plus \Ni), our derived $P_0$ ($10.42$ ms) is significantly
larger than that (1.2 ms) of B21; our {derived} $B$ ($6.52 \times 10^{14}$ G) is about 1/6
of that ($4 \times 10^{15}$ G) of B21.
{We suggest that the discrepancies between our values of $B$ and $P_0$ and those of B21 might be due to the differences
of explosion epoch and other parameters, since the large parameter space involved in the magnetar plus \Ni model
might yield another parameter sets that can fit the same LCs.
However,} our derived values of $B$ and $P_0$ are not extreme and comparable
to those of {luminous SNe} studied in the literature,
indicating that the magnetar plus \Ni model is reasonable.

{Furthermore}, our derived \Ni mass is 0.08 \Msun, which can be reduced by to {0.05 \Msun
by using Equation \ref{equ:newNimass}. This value} is much lower than the \Ni mass derived {by using the} \Ni model,
{suggesting} that the magnetar can effectively {reduce} the \Ni needed {to power} the LCs of SN~2018gk.
{More importantly}, the derived \Ni is lower than the upper limit ($\sim$ 0.2 \Msun) of the \Ni mass that
can be synthesized by neutrino-power mechanism, {and comparable to the \Ni masses of typical SNe IIb} \footnote{
{\cite{Afsariardchi2021} derive the \Ni masses of 8 SNe IIb, finding that the \Ni masses of
SN~1993J, SN~1996cb, SN~2006T, SN~2006el, SN~2008ax, SN~2011dh, SN~2013df, and SN~2016gkg are
0.08 \Msun, 0.03 \Msun, 0.08 \Msun, 0.05 \Msun, 0.06 \Msun, 0.09 \Msun, 0.06 \Msun, 0.05 \Msun, respectively
(see Table 2 of \citealt{Afsariardchi2021}). The mean and the median values of the small SNe IIb sample are 0.0625
\Msun and 0.06 \Msun, respectively.}}.

\subsection{Modeling the multi-band LCs of SN~2018gk using the Fallback Plus \Ni model}
\label{subsec:fallback+ni}

{The fallback and fallback plus \Ni models are also used to fit the multi-band LCs of some SNe
(see, e.g., the references of B21 and \citealt{Moriya2018}).
B21 disfavor the fallback model since the \Ni synthesized by the SN shock
is accreted on to the central compact remnant and very low or no \Ni mass can survive.
Here, we check the possibility that the multi-band LCs of SN~2018gk can be fitted by the fallback plus \Ni model.}

The power function $L_{\rm input}(t)$ of the fallback model is \citep{Moriya2018}.
\begin{eqnarray}
	L_{\rm input}(t) = L_{\rm input,fb}(t) =
	\left\{
	\begin{array}{lr}
		L_1\left(t_{\rm tr,fb}/1 \,\rm s\right)^{-5/3}, &\qquad t < t_{\rm tr,fb} \\
		L_1\left(t/1 \,\rm s \right)^{-5/3}, &\qquad t\geq t_{\rm tr,fb}\\
	\end{array}
	\right.
	\label{equ:input-fallback}
\end{eqnarray}
where $L_1$ is a constant and $t_{\rm tr,fb}$ is {the} transition time when the constant became
a power-law function \citep{Moriya2018}.

By combining Equations \ref{equ:lum}, \ref{equ:SED}, \ref{equ:2-v-R}, \ref{equ:2-v-T}, \ref{equ:input-Ni-positron}, and \ref{equ:input-fallback},
the fallback plus \Ni model can be constructed.
The definitions, the units, and the priors of the parameters of the fallback plus \Ni model are listed in Table \ref{table:parameters-fb}.
The MCMC using the \texttt{emcee} Python package is adopted to get the best-fitting parameters and the
1-$\sigma$ range of the parameters. We employ 20 walkers, each walker runs 30,000 steps
for the fallback plus \Ni model. The 1-$\sigma$ uncertainties are corresponding to the 16th and
84th percentiles of the posterior samples.

\begin{figure}[tbph]
\begin{center}
\includegraphics[width=0.7\textwidth,angle=0]{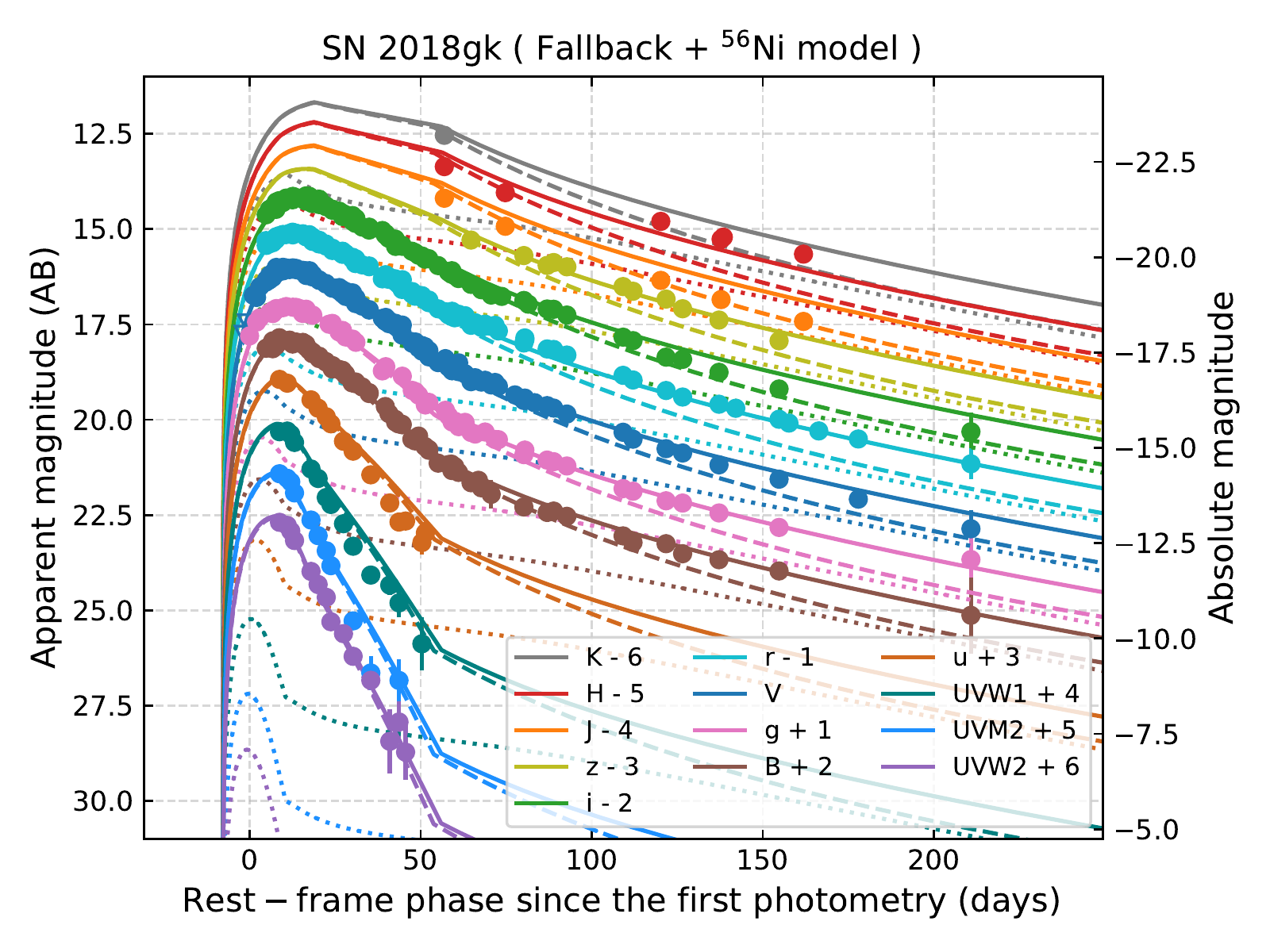}
\end{center}
\caption{The best fits (the solid curves) of the multi-band LCs
	of SN~2018gk using the fallback plus \Ni model. The shaded regions indicate 1-$\sigma$ bounds
	of the parameters. The dotted lines and dashed lines
	are the LCs powered by the \Ni and the fallback, respectively.
	The data are from Table 1 of B21, triangles represent upper limits.}
\label{fig:multibandfits-fb}
\end{figure}

The fit of the fallback plus \Ni model is shown in Figure \ref{fig:multibandfits-fb}.
The parameters and the {corresponding corner plot are} presented in
Table \ref{table:parameters-fb} and Figure \ref{fig:corner_fallback+ni}, respectively.
We find that the LCs can be well fitted by the fallback plus \Ni model, although
the LCs are rather narrow.
The relevant parameters of the fallback plus \Ni model are
{$M_{\rm ej} =$ 1.70 \Msun, $L_1 = 8.71 \times 10^{53}$ erg s$^{-1}$, $t_{\rm tr,fb} = $ 14.12 days, $M_{\rm Ni} =$ 0.05 \Msun,
$t_{\rm tr} = 19.32$ days, $v_{\rm ph1}= 0.91 \times 10^9$ cm s$^{-1}$, $v_{\rm ph2}=0.13 \times 10^9$ cm s$^{-1}$,
$T_{\rm f}=5235.48 $ K, {$t_{\rm shift} = -$8.11 days (the offset of the best-fitting explosion
epoch with respect to the explosion epoch adopted by B21 is 7.71 days),
$\kappa_{\gamma,\rm Ni} =$ 0.20 cm$^2$g$^{-1}$ ($t_{0,\gamma,\rm Ni}$ = 145.53 days),
and $\kappa_{\gamma,\rm fb} =$ 0.031 cm$^2$g$^{-1}$ ($t_{0,\gamma,\rm fb}$ = 58.60 days)}.
Using Equation \ref{equ:newNimass}, \Ni mass is reduced to 0.03 \Msun.}

Based on the derived parameters, we can inferred accretion mass $M_{acc}$ which can be expressed as \citep{Moriya2018}
\begin{eqnarray}
\eta M_{\rm acc}c^2 = \int_0^{\infty}L_{\rm input,fb}(t) dt = 2.5L_1t_{\rm tr,fb}^{-2/3}
\end{eqnarray}
here, $\eta$ is the efficiency of converting accretion to input energy.
The typical value of $\eta$ is assumed to be $\sim 0.001$ \citep{Dexter2013}.
However, $\eta$ can also be supposed to be an extreme value of
$\sim 0.1$ (e.g., \citealt{McKinney2005,Kumar2008,Gilkis2016}).
Therefore, the range of $\eta$ value can be $\sim 0.001-0.1$.
According to the derived values of $L_1$ and $t_{\rm tr,fb}$,
we find that the accretion mass of SN~2018gk is {$0.001-0.11$ \Msun.}

The problem of the fallback plus \Ni model is that a decreasing \Ni mass
needs a input function including a time-variant \Ni mass.
However, this effect can be neglected if $M_{\rm acc}$ is very small (e.g., $<0.01$\Msun).
{The accretion mass $M_{\rm acc}$ is only $0.001-0.011$ \Msun,
if $\eta$ is $0.01-0.1$; the accretion mass can reach about 0.1 \Msun,
which is larger than the initial \Ni mass (0.03 \Msun) we derived,
if $\eta$ is as low as about 0.001.
Therefore, the fallback plus \Ni model favor a relatively large $\eta$,
since smaller $\eta$ would result in a large accreted \Ni mass higher than
the \Ni mass derived by the model. Therefore, although the fallback plus \Ni
model cannot be excluded, its validity depends on the value of $\eta$.}

\section{The NIR Excesses of SN~2018gk And the properties of Possible Dust}
\label{sec:dust}

\begin{figure}[tbph]
	\begin{center}
		\includegraphics[width=0.32\textwidth,angle=0]{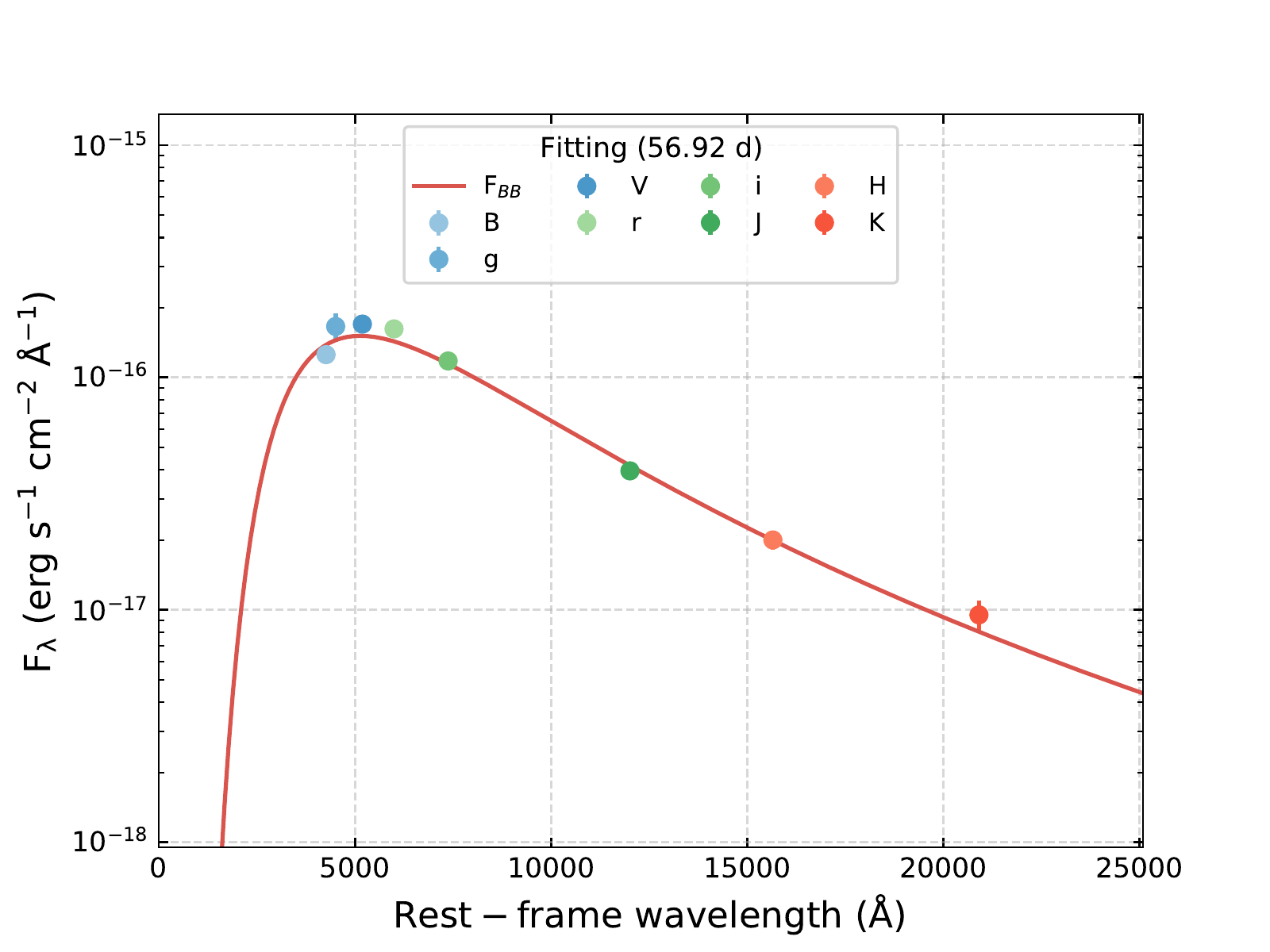}
		\includegraphics[width=0.32\textwidth,angle=0]{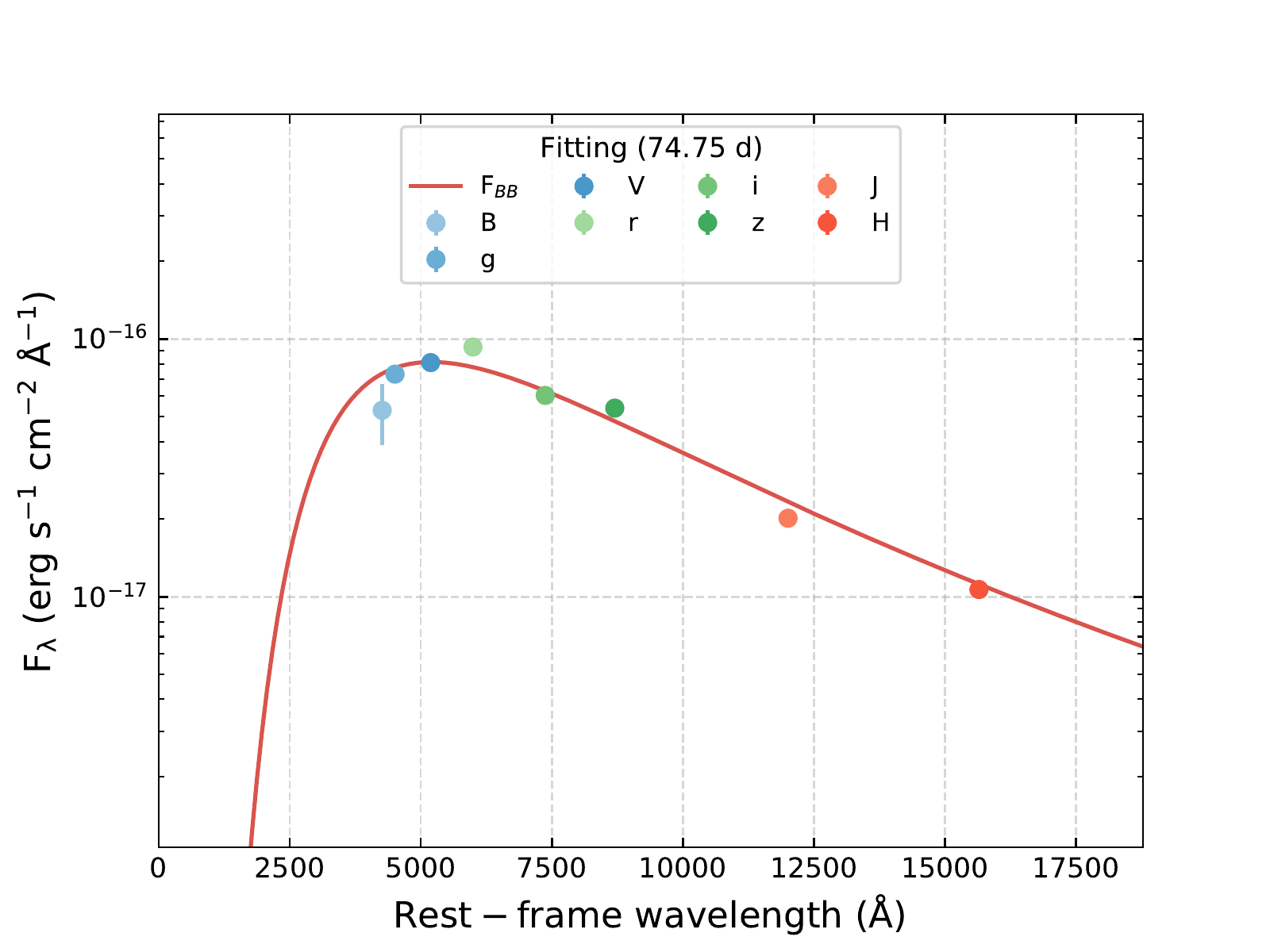}
		\includegraphics[width=0.32\textwidth,angle=0]{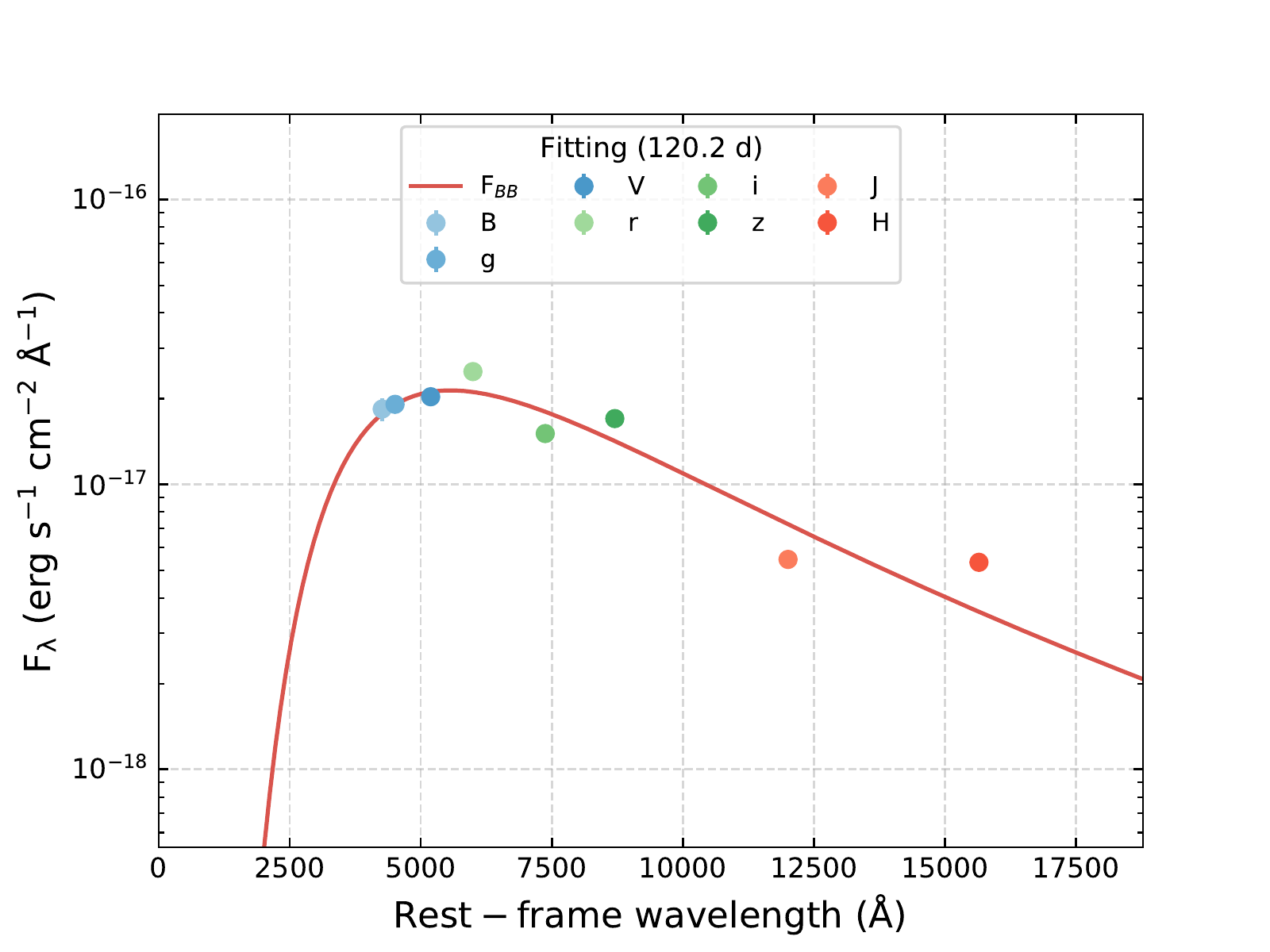}
		\includegraphics[width=0.32\textwidth,angle=0]{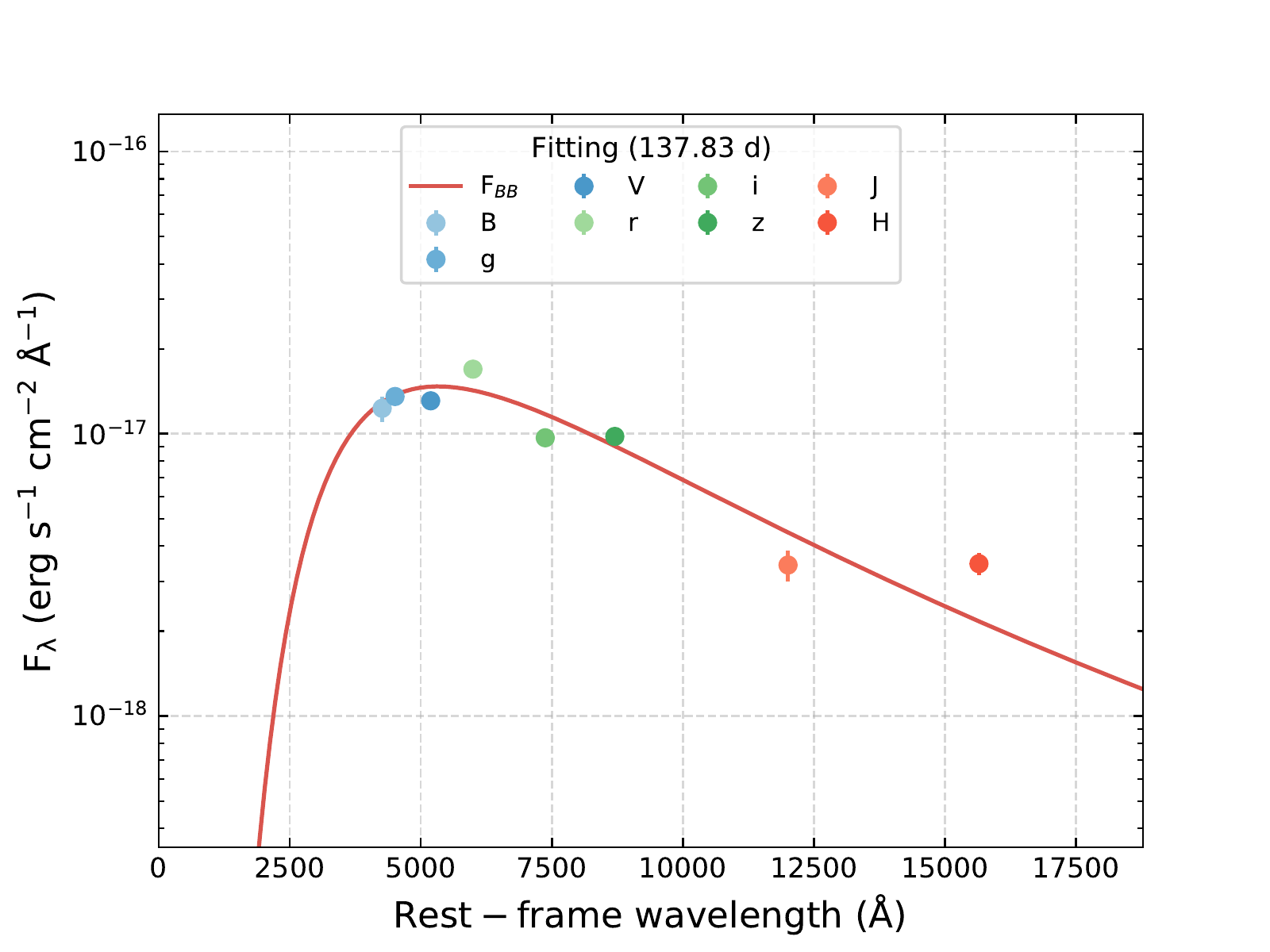}
		\includegraphics[width=0.32\textwidth,angle=0]{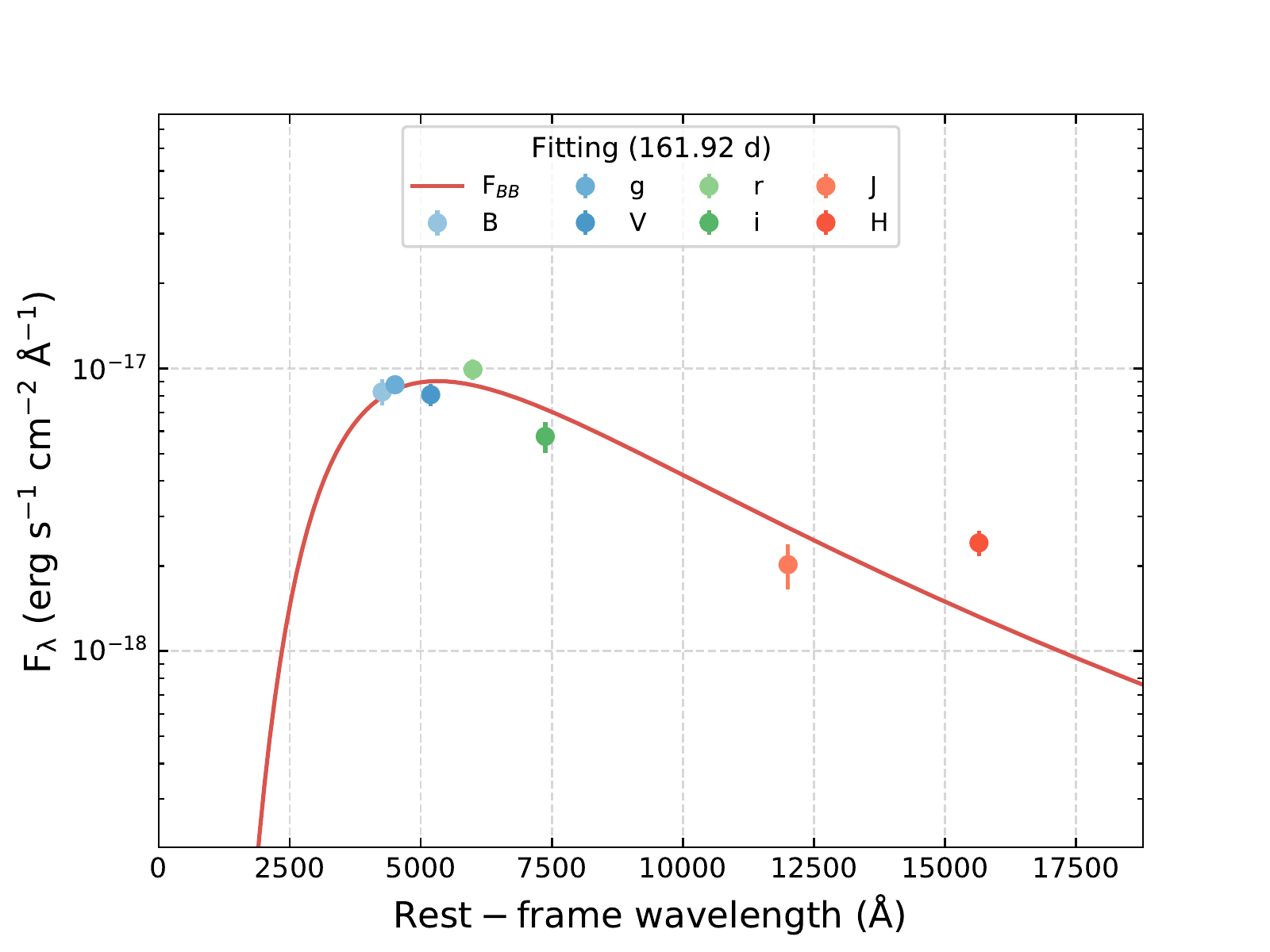}
	\end{center}
	\caption{The blackbody fits for SEDs of SN~2018gk at five epochs.}
	\label{fig:blackbod}
\end{figure}

As pointed out by B21, SN~2018gk show NIR excesses in the $H-$band flux,
which can be due to the strong emission in the $H-$band or the dust emission.
Our multi-band fits also show the $H-$band excesses at the late-epochs.
Furthermore, our blackbody fits for the optical$-$NIR SEDs at five epochs when at
least $J-$ and $H-$ band photometry are available demonstrate that, while
the SEDs at first two epochs do not show NIR excesses, the
SEDs at the rest three epochs show evident NIR excesses (see Figure \ref{fig:blackbod},
{the derived parameters are listed in Table \ref{table:blackbody}}).

\begin{figure}[tbph]
	\begin{center}
		\includegraphics[width=0.48\textwidth,angle=0]{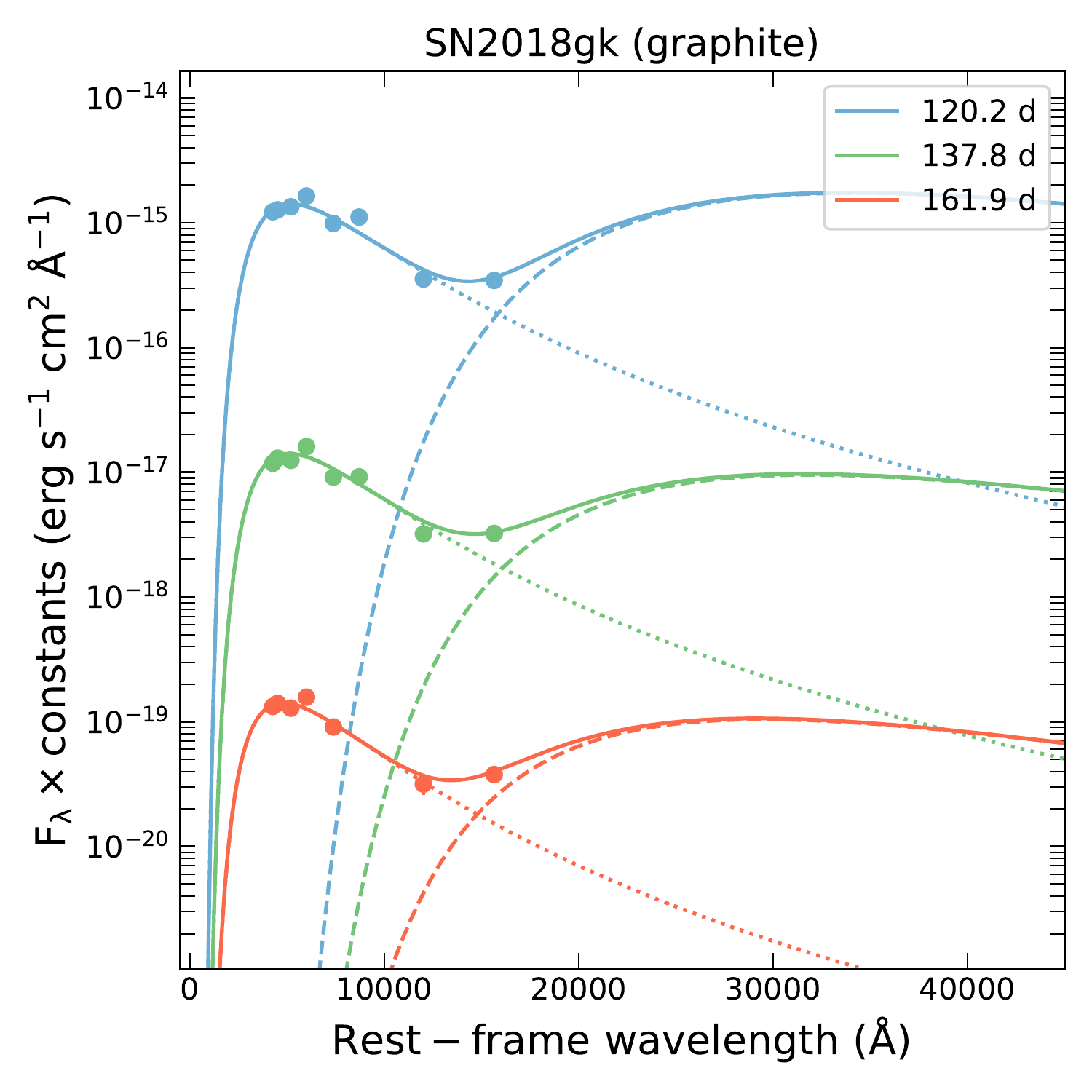}
		\includegraphics[width=0.48\textwidth,angle=0]{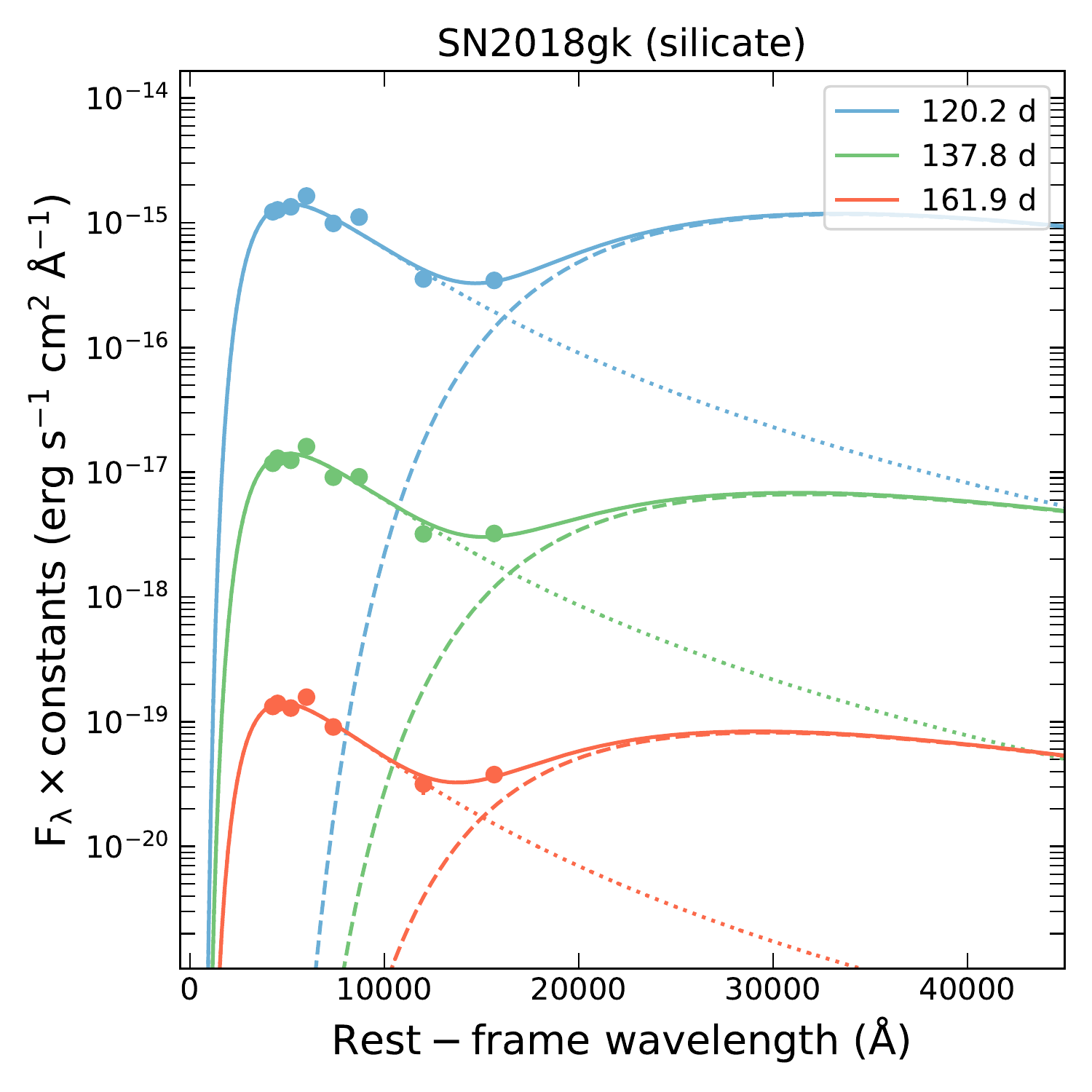}
	\end{center}
	\caption{The blackbody plus dust emission model fits for SEDs of SN~2018gk at three epochs.
The graphite (left panel) and silicate (right panel) dust are adopted.
The dotted, dashed, and solid lines represent the flux of the photospheres, the dust, and the sum of the two components, respectively.}
	\label{fig:dust}
\end{figure}

We suppose that the NIR excesses of the SEDs are due to dust emission, and
adopt a blackbody plus dust emission model to fit the optical$-$NIR SEDs of SN~2018gk.
The details of the model can be found in \cite{Gan2021}, and we suppose that the
dust size distribution follow a power law function (see, e.g., \citealt{Cao2022} and the
references therein). The MCMC method is also used.
The free parameters of the model are the dust
mass $M_d$, the dust temperature $T_d$, the temperature of the SN photosphere $T_{\rm ph}$, the
radius of the SN photosphere $R_{\rm ph}$.
We suppose that the dust might be graphite or silicate.
The fits of the model are shown in Figure \ref{fig:dust},
the derived parameters are listed in Table \ref{table:dust}.

We find that the putative dust mass is $\sim 2.9-19.4 \times 10^{-3}$ \Msun or $\sim 3.5-24.9 \times 10^{-3}$ \Msun
for graphitic or silicate dust, the temperature of the dust is $\sim$ $730-850 $ or $\sim$ $780-880$ K for
the two cases. The derived median values of the dust mass are in the range (a few $10^{-5}$ to a few $10^{-2}$ \Msun)
of those of the dust of many SNe (see, e.g., Table 5 of \citealt{Fox2013}).
Moreover, The derived temperature are lower than the
evaporation temperatures of silicate ($\sim1100-1500$ K, \citealt{Laor1993,Mattila2008,Gall2014})
and graphite ($\sim 1900$ K, \citealt{Stritzinger2012}).
These two facts support the assumption that
the NIR excesses are from the dust emission.

However, the absence of data in $K-$ band and other bands with longer effective wavelengths at the three epochs
prevents us from posing more stringent constraints on the temperature
and mass of the putative dust.
Especially, the absence of data in bands with longer effective wavelengths
would result in a longer derived peak wavelength
of the SEDs of the dust emission. A temperature lower than the real value
is therefore favorable. Hence, the dust
temperature of SN~2018gk we derive can be regarded as the lower limit
of the real temperature of the putative dust.

\section{Discussion}
\label{sec:discussion}

\subsection{The Mass of the Ejecta}

The ejecta mass of SN~2018gk derived by the magnetar plus \Ni model
we adopt is $\sim$ {1.65} \Msun.
According to the analysis of B21, the $M_{\rm ZAMS}$
and the helium core mass just prior to the explosion are $\sim 20 - 25$ \Msun and $\sim 6-8$ \Msun.
After removing the putative magnetar with mass being 1.4 \Msun, the ejecta mass is
$\sim 4.6 - 6.6$ \Msun, which is larger than our derived ejecta mass.
It seems that the discrepancy between the two value disfavor both
the two models.

The contradiction can be eliminated since (1) the magnetar mass can be up to
$\sim 2.0-2.5$ \Msun, then the the lower limit of the ejecta can be reduced
to 3.5 \Msun; (2) the final masses of the progenitors of SESNe are sensitive
to metallicity and the $M_{\rm ZAMS}$, the SESNe progenitor with
$M_{\rm ZAMS} \sim 18 - 25$ \Msun might yield ejecta masses of $2.39 - 3.61$ \Msun
\citep{Dessart2011}; (3) the analysis of B21 is based on the
single-star evolution model, while the binary-star evolution can
reduce the helium core masses of the progenitor masses of SESNe \citep{Yoon2010};
(4) a smaller $\kappa$ would result in a larger ejecta mass.

Hence, we suggest that the ejecta mass of SN~2018gk derived by the magnetar
plus \Ni model and the model itself are reasonable. The discrepancy between
the two values might indicate that the progenitor of SN~2018gk was in a binary system
and/or had a low metellicity, or the explosion of SN~2018gk leaved a massive
magnetar.

\subsection{The Theoretical Bolometric Light Curve, the Temperature Evolution, and the Radius Evolution of SN~2018gk}

\begin{figure}[tbph]
\begin{center}
\includegraphics[width=0.48\textwidth,angle=0]{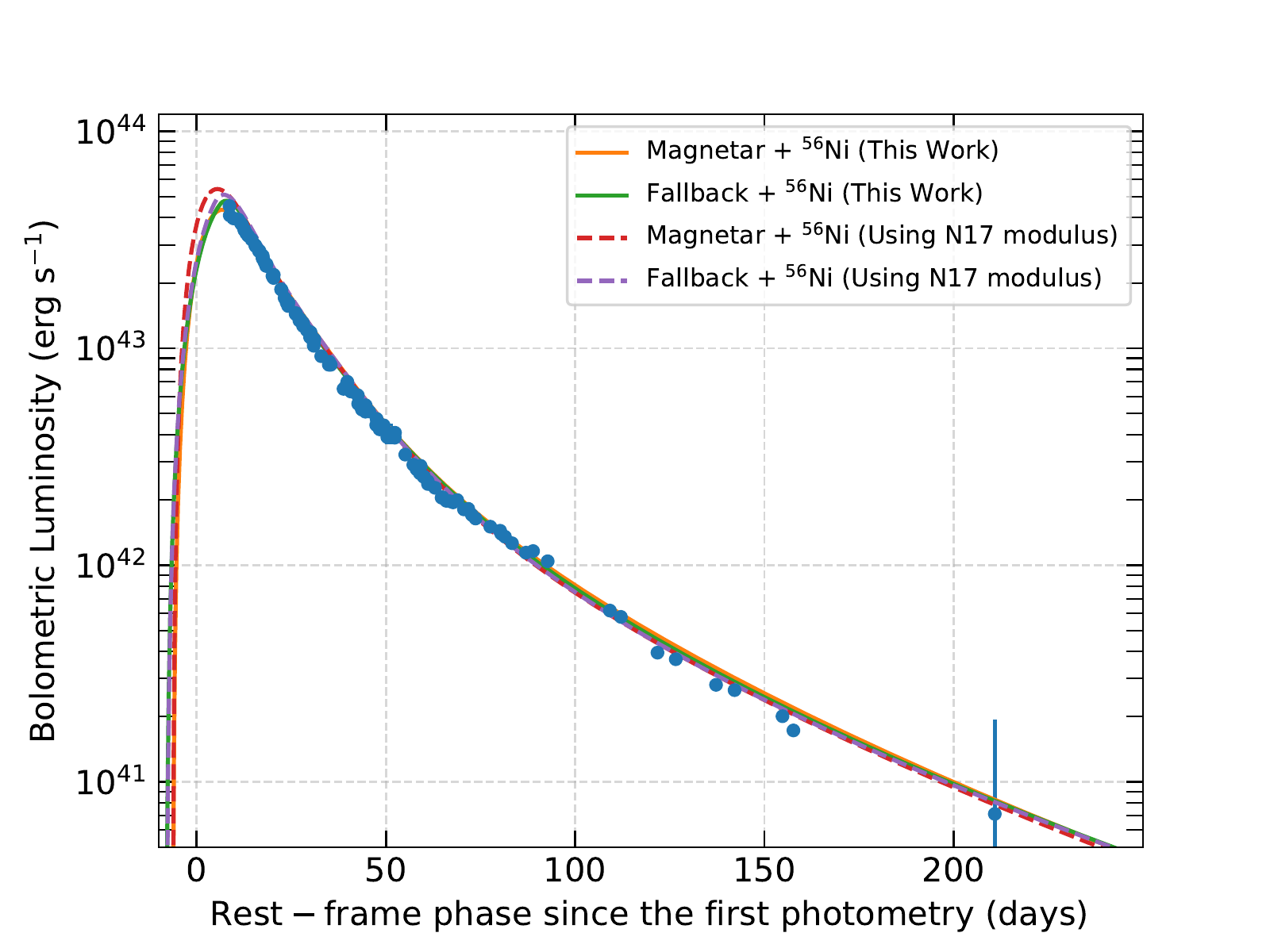}
\includegraphics[width=0.48\textwidth,angle=0]{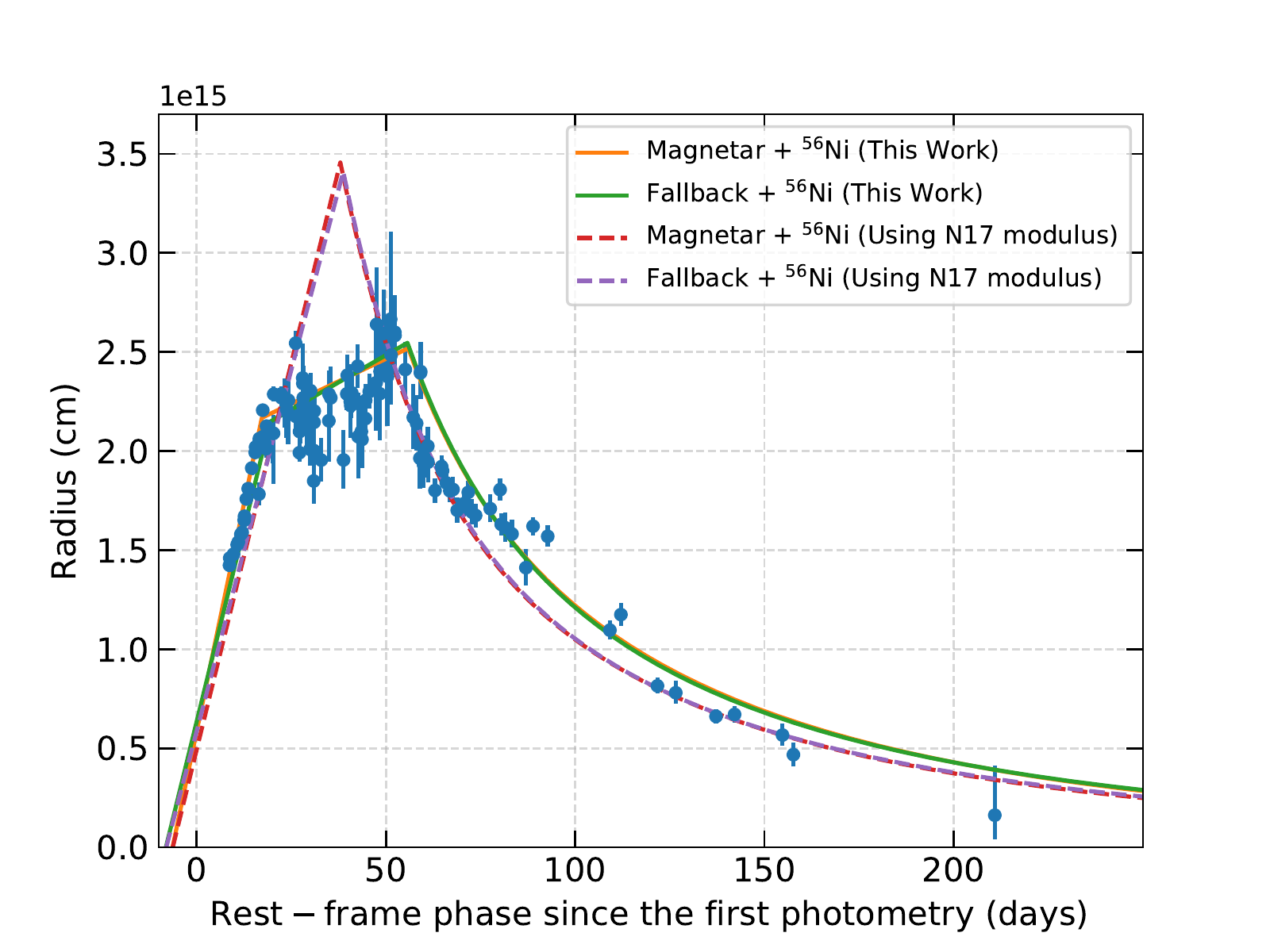}
\includegraphics[width=0.48\textwidth,angle=0]{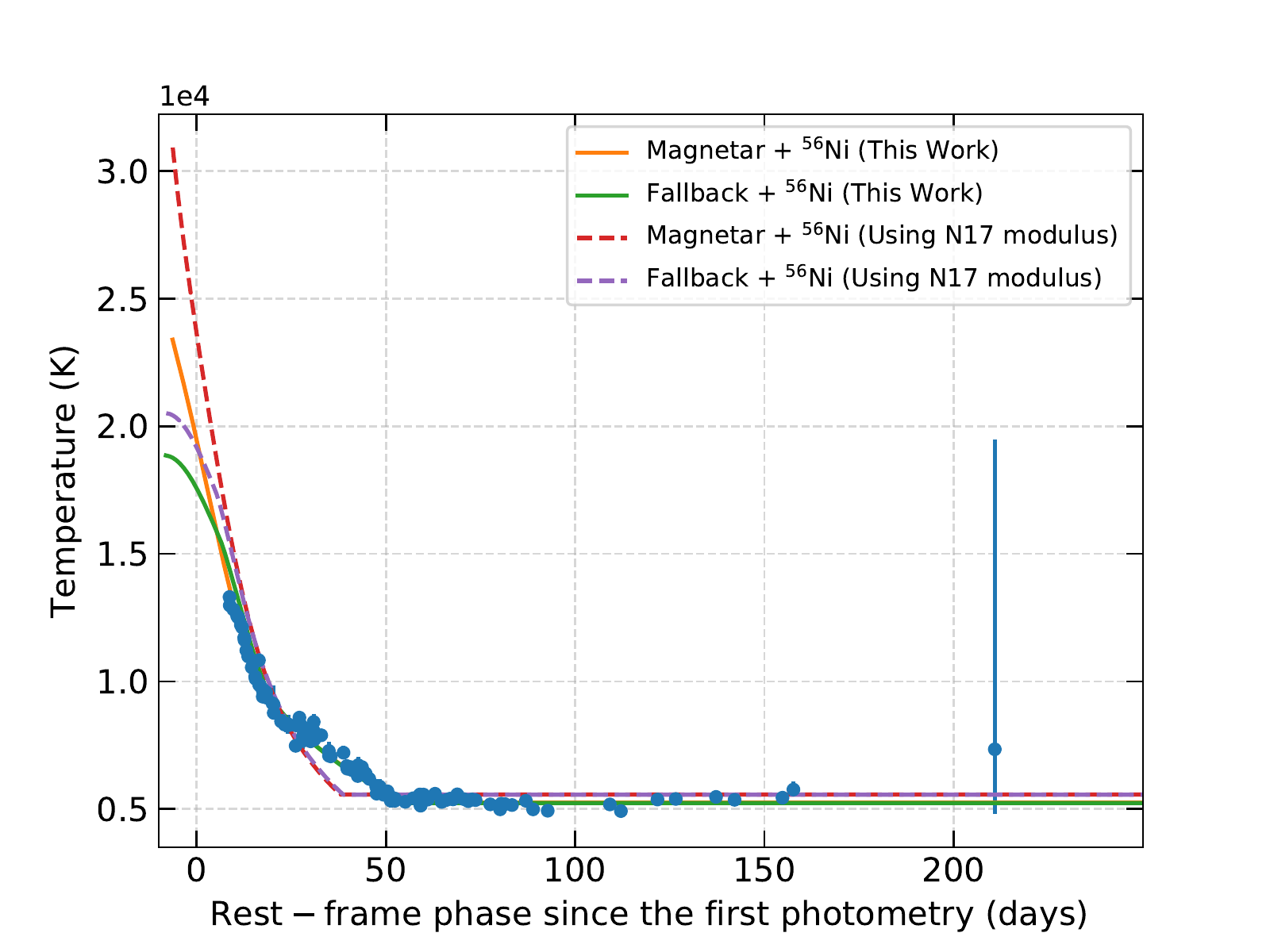}
\end{center}
\caption{The bolometric LCs (top-left panel), the radius evolution (top-right panel), and the temperature
	evolution (the bottom panel) reproduced by the best-fitting parameters of the magnetar plus \Ni
	model ({orange} solid lines) and fallback plus \Ni model ({green} solid lines), and
{the bolometric LCs, the radius evolution, and the temperature evolution} derived
 from the observations ({blue} filled points).
For comparison, the bolometric LCs, radius evolution, and temperature evolution using the photosphere modulus
{of} \cite{Nicholl2017} (N17) are also presented {(red and purple dashed lines).}
{The $t_{\rm shift}$ values of our magnetar plus \Ni and fallback plus \Ni models are $-$6.35 days and $-$8.11 days, respectively;
the $t_{\rm shift}$ values of the magnetar plus \Ni and fallback plus \Ni model using N17's the photosphere modulus are $-$6.30 days and $-$8.01 days, respectively}.}
\label{fig:bolo}
\end{figure}

Based on the best-fitting parameters of the magnetar plus \Ni
model obtained from the multi-band LCs, the theoretical
bolometric LC, temperature evolution, and the
radius evolution can be reproduced, see Figure \ref{fig:bolo}.
For comparison, the same three curves reproduced by the
best-fitting parameters of the model adopting photosphere modulus
of \cite{Nicholl2017} are also plotted in the same figure.

We find that while the theoretical bolometric LC of both models can fit the
the bolometric LC derived by the photometry, temperature and the radius evolution curves
reproduced by our model which use a generalized photosphere modulus can better match those derived
by the photometry. This indicate the photosphere velocity have decreased before the temperature
reached the temperature floor ($T_{\rm f}$), and demonstrate the necessity of generalizing the
photosphere modulus.

We find the rise time of the bolometric LC of SN~2018gk {reproduced by the best-fitting
parameters of the magnetar plus \Ni model}
is $\sim$ 13.2 days which is significantly {longer than} the rise time
($\lesssim 5$ days, see Figure 14 of B21) of the bolometric LC reproduced by B21 using
the magnetar plus \Ni model. The peak luminosity of the bolometric LC {reproduced}
is $\sim 4.35 \times 10^{43}$ erg s$^{-1}$,
which is half of the peak luminosity ($\sim$ 10$^{44}$ erg s$^{-1}$) derived by B21 using the
magnetar plus \Ni model.
{Our derived peak luminosity of the bolometric LC of SN~2018gk} also suggests that
SN~2018gk is one of luminous SNe which bridges the gap between SLSNe
(whose peak bolometric luminosity are $\gtrsim 7 \times 10^{43}$ erg s$^{-1}$, \citealt{Gal2012})
and normal-luminosity SNe.

\section{Conclusions}
\label{sec:conclusions}

SN~2018gk is a luminous SN IIb whose peak luminosity is between those of SLSNe and normal SNe.
In this paper, we perform a comprehensive study for the physical properties of SN~2018gk.
Based on the SED fits, we find that the early-time photospheric velocity
vary from a larger value to a smaller value before the photosphere reach
a temperature floor. We generalize the photosphere modulus
by assuming that the the photospheric velocity of SN~2018gk has two
different values at two time episodes before reaching the photosphere reach
the temperature floor.

By incorporating the generalized photosphere modulus to different
energy source models, we fit the multi-band LCs of SN~2018gk.
We {conclude} the \Ni model cannot fit the multi-band LCs of SN~2018gk,
since the derived \Ni mass ($\sim$0.9 \Msun) is significantly larger than
the ejecta mass ($\sim$0.10 \Msun) if the \Ni model is adopted.

In contrast, we find that the magnetar plus \Ni and the fallback
plus \Ni models can fit the multi-band LCs of SN~2018gk. It should be
pointed out that, however, {the validity of the fallback plus \Ni model
depends on the value of $\eta$, and favors a relatively large $\eta$.
Hence}, we favor the magnetar plus \Ni model, though the fallback plus \Ni model
cannot be excluded.

The \Ni mass {(0.05 \Msun)} derived here by using the magnetar plus \Ni model is significantly
lower than that ($\sim$0.9 \Msun) required by the \Ni model adopted here,
indicating that the magnetar can significantly reduce \Ni mass {required
to power the LCs of SN~2018gk}. The derived \Ni mass of the magnetar plus \Ni model
is about {1/6} of the value ($\sim$ 0.3 \Msun) derived by B21
using the same scenario, {and comparable to the \Ni masses of typical SNe IIb}.
{Moreover}, our derived \Ni mass is lower than
the upper limit (0.2 \Msun) of the value of the \Ni mass can by synthesized by the
neutrino-powered CCSNe.

The value of $P_0$ (10.42 ms) {derived by the magnetar plus \Ni model we adopt}
is about 8 times that (1.2 ms) derived by B21; the derived value of $B$
($6.52 \times 10^{14}$ G) is about 1/6 of that
($4 \times 10^{15}$ G) derived by B21. A larger $P_0$ usually
results in a lower input power and a lower peak luminosity.
In general, magnetars with $P_0 \sim 1-5$ ms and $B \sim 10^{14}-10^{15}$ G
would power SLSNe (see, e.g., \citealt{Nicholl2017,Liu2017}), while magnetar
with $P_0 \sim 5-20$ ms and $B \sim 10^{13}-10^{15}$ G would
power luminous SNe (see, e.g., \citealt{Wang2015b,Gomez2022}) with rare exceptions.
Hence, we suggest our derived parameters of the magnetar supposed to
power the LCs of SN~2018gk are reasonable, since SN~2018gk
is a luminous SN rather than a SLSN.

Therefore, we suggest that the magnetar plus a moderate amount of \Ni can
be responsible for the luminous LCs of SN~2018gk.
To our knowledge, other luminous SNe IIb requiring
central engines (plus \Ni) to account for their LCs are PTF10hgi and DES14X2fna.
So SN~2018gk might be one of rare SNe IIb mainly powered by a central engine.

Based on the best-fitting parameters of the magnetar plus \Ni model, we plot the bolometric LC,
the photospheric temperature evolution, and the photospheric radius evolution
of SN~2018gk, and compare them with those from the SED fits.
The peak of the theoretical bolometric LC of SN~2018gk is
{$\sim 4.35 \times 10^{43}$ erg s$^{-1}$} which is below the fiducial threshold of SLSNe
($\lesssim 7 \times 10^{43}$ erg s$^{-1}$; \cite{Gal2012}) and the
peak luminosity ($\sim 10^{44}$ erg s$^{-1}$ derived by B21.
The rise time of the theoretical bolometric LC of SN~2018gk is $\sim$ {13.2 days} which
is comparable to those of many SESNe.

Moreover, we find that the theoretical evolution of the temperature and the radius of the
photosphere are well matched by those from SED fits. For comparison, we also plot the
the theoretical evolution of the temperature and the radius of the photosphere based on
the simple photosphere modulus that assume the photosphere velocity are constants before
the temperature floor are reached deviated from those from the SED fits from $30-60$ after
the derived explosion date. We suggests that the conventional photosphere modulus cannot
describe the early-time photospheric evolution of a fraction of SNe and generalized
photosphere modulus would be considered as alternative ones.

Finally, we find that the blackbody plus dust emission model can account for the
NIR excesses of the SEDs of SN~2018gk at some epochs.
The derived masses of the dust are
$\sim 2.9-19.4 \times 10^{-3}$ \Msun ($\sim 3.5-24.9 \times 10^{-3}$ \Msun)
for graphite (silicate) grains, respectively; the derived temperatures of the dust of the SN
are $\sim$ $730-850 $ ($\sim$ $780-880$) K for graphite (silicate) grains, respectively.
Due to the absence of photometry at $K-$ band and other bands with longer effective wavelengths,
the real parameters cannot be well constrained, and the derived dust temperature
might be the lower limit of the real temperature of the dust.

\begin{acknowledgments}
{We thank the anonymous referee for helpful comments and suggestions that have allowed us to improve this manuscript.}
Wang Tao thank Jia-Wei Li \begin{CJK*}{UTF8}{gbsn}(李佳伟)\end{CJK*}, Shuai-Cong Wang \begin{CJK*}{UTF8}{gbsn}(王帅聪)\end{CJK*},
Deng-Wang Shi \begin{CJK*}{UTF8}{gbsn}(石登旺)\end{CJK*}, Song-Yao Bai \begin{CJK*}{UTF8}{gbsn}(白松瑶)\end{CJK*},
and Jing-Yao Li \begin{CJK*}{UTF8}{gbsn}(李京谣)\end{CJK*} for helpful discussion.
This work is supported by National Natural Science Foundation of China (grant Nos. 11963001, 12133003).

\end{acknowledgments}

\clearpage

\begin{deluxetable}{cccccc}
	\tablecaption{The Definitions, the units, the prior, the medians, 1-$\sigma$ bounds, and the best-fitting values for the parameters of the \Ni model. The values of $\chi^{\rm 2}$/dof (reduced $\chi^{\rm 2}$, dof=degree of freedom) are also presented.}
	\label{table:parameters-ni}
	\tablehead{\colhead{Parameters}	&\colhead{Definition}  								&  \colhead{Unit}  &   \colhead{Prior}   	& \colhead{Best fit} & \colhead{Median}}
	\startdata
	$M_{\rm ej}$                    & The ejecta mass                                   &   M$_\odot$          &    $[0.1, 50]$     	&0.10 	&	$	0.10 	^{+	0.00 	}_{-	0.00 	}$ \\
	$t_{\rm tr}$                    & The photospheric velocity transition time         &   day			       &    $[0, 100]$    		&14.02 	&	$	14.03	^{+	0.11 	}_{-	0.11 	}$ \\
	$v_{\rm ph1}$                   & The first {episode} photospheric velocity  &   $10^9$ cm s$^{-1}$ &    $[0, 5]$    		&1.52 	&	$	1.52 	^{+	0.01 	}_{-	0.01 	}$ \\
	$v_{\rm ph2}$                   & The second {episode} photospheric velocity &   $10^9$ cm s$^{-1}$ &    $[0, 5]$    		&0.16 	&	$	0.16 	^{+	0.01 	}_{-	0.01 	}$ \\
	$M_{\rm Ni}$                    & The \Ni mass                                		&   M$_\odot$          &    $[0.0, 4]$    		&1.59 	&	$	1.58 	^{+	0.01 	}_{-	0.01 	}$ \\
	$\log \kappa_{\rm \gamma, Ni}$  & {The $\gamma$-ray opacity of \Ni-cascade-decay}&   cm$^2$g$^{-1}$     &    $[-1.568, 4] $      &-0.45 	&	$	-0.45 	^{+	0.01 	}_{-	0.01 	}$ \\
	$\kappa_{\rm e^{+}}$  			& {The positron opacity}						&   cm$^2$g$^{-1}$     &    $[4, 14] $      	&4.00 	&	$	4.00	^{+	0.01	}_{-	0.00 	}$ \\
	$T_{\rm f}$                     & The temperature floor of the photosphere          &   K                  &    $[1000, 10^4] $ 	&4921.90&	$	4920.83	^{+	15.20 	}_{-	15.00	}$ \\	
	$t_{\rm shift}$                 & The explosion time relative to the first data     &   day                &    $[-20, 0]$      	&-1.76 	&	$	-1.78 	^{+	0.05 	}_{-	0.05 	}$ \\
	$\lambda_{\rm CF}$		        & The cutoff wavelength							    &   {\AA}              &    $[0, 4000]$     	&2005.60&	$	2014.99 ^{+	15.44 	}_{-	10.22 	}$ \\
	$\beta'$                 		& The dimensionless free parameter       			&   		           &    $[0, 10]$       	&4.47 	&	$	3.61 	^{+	4.21 	}_{-	3.40 	}$ \\
	$\chi^{\rm 2}$/dof              &                                                  	&                      &                    	&9.30 	& 9.30 \\
	\enddata
\end{deluxetable}

\clearpage

\begin{deluxetable}{cccccc}
	\tablecaption{The Definitions, the units, the prior, the medians, 1-$\sigma$ bounds, and the best-fitting values for the parameters of the magnetar plus \Ni model. The values of $\chi^{\rm 2}$/dof (reduced $\chi^{\rm 2}$, dof=degree of freedom) are also presented.}
	\label{table:parameters-mag}
	\tablehead{\colhead{Parameters}	&\colhead{Definition}  								&  \colhead{Unit}  &   \colhead{Prior}   	& \colhead{Best fit} & \colhead{Median}}
	\startdata
	$M_{\rm ej}$                    & The ejecta mass                                   &   M$_\odot$          &    $[0.1, 50]$     	&1.65 	&	$	1.61 	^{+	0.08 	}_{-	0.08 	}$ \\
	$P_0$                           & The initial period of the magnetar                &   ms                 &    $[0.8, 50]$     	&10.42 	&	$	10.43	^{+	0.03 	}_{-	0.03 	}$ \\
	$B$                             & The magnetic field strength of the magnetar       &   $10^{14}$ G        &    $[0.1, 100]$    	&6.52 	&	$	6.48 	^{+	0.11 	}_{-	0.11 	}$ \\
	$t_{\rm tr}$                    & The photospheric velocity transition time         &   day			       &    $[0, 100]$    		&17.15 	&	$	17.01 	^{+	0.11 	}_{-	0.11 	}$ \\
	$v_{\rm ph1}$                   & The first {episode} photospheric velocity  &   $10^9$ cm s$^{-1}$ &    $[0, 5]$	    	&1.07 	&	$	1.07 	^{+	0.01 	}_{-	0.01 	}$ \\
	$v_{\rm ph2}$                   & The second {episode} photospheric velocity &   $10^9$ cm s$^{-1}$ &    $[0, 5]$			&0.11 	&	$	0.11 	^{+	0.01 	}_{-	0.01 	}$ \\
	$M_{\rm Ni}$                    & The \Ni mass                                		&   M$_\odot$          &    $[0, 0.2M_{\rm ej}]$&0.08 	&	$	0.08 	^{+	0.00 	}_{-	0.00 	}$ \\
	$\log \kappa_{\rm \gamma, mag}$ & {The $\gamma$}-ray opacity of magnetar photons         &   cm$^2$g$^{-1}$     &    $[-2, 4] $      	&-1.55 	&	$	-1.55 	^{+	0.02 	}_{-	0.02 	}$ \\
	$\log \kappa_{\rm \gamma, Ni}$  & {The $\gamma$-ray opacity of \Ni-cascade-decay photons}&   cm$^2$g$^{-1}$     &    $[-1.568, 4] $      &-0.72 	&	$	-0.72 	^{+	0.04 	}_{-	0.04 	}$ \\
	$\kappa_{\rm e^{+}}$  			& {The positron opacity}						&   cm$^2$g$^{-1}$     &    $[4, 14] $      	&7.02	&	$	8.95	^{+	3.43	}_{-	3.35  	}$ \\	
	$T_{\rm f}$                     & The temperature floor of the photosphere          &   K                  &    $[1000, 10^4] $ 	&5254.04&	$	5244.27 ^{+	17.10 	}_{-	17.11	}$ \\	
	$t_{\rm shift}$                 & The explosion time relative to the first data     &   day                &    $[-20, 0]$      	&-6.35 	&	$	-6.29 	^{+	0.07 	}_{-	0.06 	}$ \\	
	$\lambda_{\rm CF}$		        & The cutoff wavelength							    &   {\AA}              &    $[0, 4000]$     	&2007.99&	$	2019.53 ^{+ 1302.88 }_{-	13.33 	}$ \\	
	$\beta'$                 		& The dimensionless free parameter       			&   		           &    $[0, 10]$       	&3.82 	&	$	1.95 	^{+	5.33	}_{-	1.92 	}$ \\
	$\chi^{\rm 2}$/dof              &                                                  	&                      &                    	&6.08 	& 6.08 \\
	\enddata
\end{deluxetable}

\clearpage

\begin{deluxetable}{cccccc}
	\tablecaption{The Definitions, the units, the prior, the medians, 1-$\sigma$ bounds, and the best-fitting values for the parameters of the fallback plus \Ni model. The values of $\chi^{\rm 2}$/dof (reduced $\chi^{\rm 2}$, dof=degree of freedom) are also presented.}
	\label{table:parameters-fb}
	\tablehead{\colhead{Parameters}	&\colhead{Definition}  								&  \colhead{Unit}  &   \colhead{Prior}   	& \colhead{Best fit} & \colhead{Median}}
	\startdata
	$M_{\rm ej}$                    & The ejecta mass                                  		 	&   M$_\odot$          &    $[0.1, 50]$     	&1.70 	&	$	1.66 	^{+	0.15 	}_{-	0.17 	}$ \\
	$\log L_1$                      & The constant        										&   erg s$^{-1}$       &    $[48, 58]$     		&53.94 	&	$	53.94 	^{+	0.00 	}_{-	0.01 	}$ \\
	$\log t_{\rm tr,fb}$          	& The transition time										&   day        		   &    $[-3, 3]$    		&1.15 	&	$	1.15 	^{+	0.01 	}_{-	0.01 	}$ \\
	$t_{\rm tr}$                    & The photospheric velocity transition time        	 		&   day			       &    $[0, 100]$    		&19.32 	&	$	19.04 	^{+	0.32 	}_{-	0.45 	}$ \\
	$v_{\rm ph1}$                   & The first {episode} photospheric velocity  				&   $10^9$ cm s$^{-1}$ &    $[0, 5]$    		&0.91 	&	$	0.91 	^{+	0.01 	}_{-	0.01 	}$ \\
	$v_{\rm ph2}$                   & The second {episode} photospheric velocity 			 	&   $10^9$ cm s$^{-1}$ &    $[0, 5]$      		&0.13 	&	$	0.14 	^{+	0.02 	}_{-	0.02 	}$ \\
	$M_{\rm Ni}$                    & The \Ni mass                                			 	&   M$_\odot$          &    $[0, 0.2M_{\rm ej}]$&0.05 	&	$	0.05 	^{+	0.00 	}_{-	0.00 	}$ \\
	$\log \kappa_{\gamma, \rm fb}$  & {The $\gamma$}-ray opacity of fallback photons    	 	&   cm$^2$g$^{-1}$     &    $[-2, 4] $      	&-1.58 	&	$	-1.56 	^{+	0.04 	}_{-	0.03 	}$ \\
	$\log \kappa_{\gamma, \rm Ni}$  & {The $\gamma$-ray opacity of \Ni-cascade-decay photons}	&   cm$^2$g$^{-1}$     &    $[-1.568, 4] $      &-0.79 	&	$	-0.75 	^{+	0.06 	}_{-	0.07 	}$ \\
	$\kappa_{\rm e^{+}}$  			& {The positron opacity}									&   cm$^2$g$^{-1}$     &    $[4, 14] $      	&12.11 	&	$	9.00	^{+	3.43	}_{-	3.45 	}$ \\
	$T_{\rm f}$                     & The temperature floor of the photosphere          		&   K                  &    $[1000, 10^4] $ 	&5235.48&	$	5227.37 ^{+	17.13 	}_{-	17.03 	}$ \\	
	$t_{\rm shift}$                 & The explosion time relative to the first data     		&   day                &    $[-20, 0]$      	&-8.11 	&	$	-8.07 	^{+	0.13 	}_{-	0.10 	}$ \\	
	$\lambda_{\rm CF}$		        & The cutoff wavelength							    		&   {\AA}              &    $[0, 4000]$     	&3995.45&	$	3980.84 ^{+	14.25 	}_{-	28.87 	}$ \\	
	$\beta'$                 		& The dimensionless free parameter       			 		&   		           &    $[0, 10]$       	&0.30 	&	$	0.30 	^{+	0.03 	}_{-	0.03 	}$ \\	
	$\chi^{\rm 2}$/dof              &                                                  			&                      &                    	&6.38	& 6.39 \\
	\enddata
\end{deluxetable}

\clearpage

\begin{table*}
	\setlength{\tabcolsep}{18pt}
	\renewcommand{\arraystretch}{1.5}
	\caption{The best-fitting parameters (the temperature of the photosphere $T_{\rm ph}$, the radius of the photosphere $R_{\rm ph}$) of the blackbody model for the SEDs of SN~2018gk.}
	\label{table:blackbody}
	\centering
	\begin{tabular}{cccc}
		\hline\hline
		\colhead{Phase\footnote{All the phases are relative to the date of the first data which is MJD 58131. All epochs are transformed to the rest-frame ones.}} & \colhead{$T_{\rm ph}$}&\colhead{$R_{\rm ph}$} & \colhead{$\chi^{\rm 2}$/dof} \\
		(days) & (10$^{3}$\,K) & (10$^{15}$ cm)    & \\\hline
		56.92 	&$	5689.72 	^{+	100.62 	}_{-	93.12 	}$&$	1.95 	^{+	0.07 	}_{-	0.07 	}$&	1.51 	\\\hline
		74.75 	&$	5604.40 	^{+	64.62 	}_{-	63.81 	}$&$	1.48 	^{+	0.03 	}_{-	0.03 	}$&	4.82 	\\\hline
		120.20 	&$	5238.48 	^{+	83.88 	}_{-	81.47 	}$&$	0.90 	^{+	0.03 	}_{-	0.03 	}$&	12.33 	\\\hline
		137.83 	&$	5479.72 	^{+	126.27 	}_{-	114.47 	}$&$	0.67 	^{+	0.03 	}_{-	0.03 	}$&	9.12 	\\\hline
		161.92 	&$	5477.06 	^{+	182.17 	}_{-	164.62 	}$&$	0.52 	^{+	0.04 	}_{-	0.04 	}$&	4.44 	\\	
		\hline
		\hline	
	\end{tabular}
\end{table*}

\clearpage

\begin{table*}
	\centering
	\footnotesize
	\setlength{\tabcolsep}{1.8pt}
	\renewcommand{\arraystretch}{1.5}
	\caption{The best-fitting parameters (the temperature of the photosphere $T_{\rm ph}$, the radius of the photosphere $R_{\rm ph}$, the temperature of the dust shell $T_{\rm d}$, and the mass of the dust shell $M_{\rm d}$) of the blackbody plus graphite (silicate) model for the SEDs of SN~2018gk.}
	\label{table:dust}
	\centering
	\begin{tabular}{c c c c c c c c c c c c c}
		\hline\hline\noalign{\smallskip}
		\multicolumn{5}{c}{{Blackbody plus Graphite}} & \multicolumn{5}{c}{{Blackbody plus Silicate}}\\\noalign{\smallskip}
		\cmidrule(lr){2-6} \cmidrule(lr){7-11}
		\colhead{Phase\footnote{All the phases are relative to the date of the first data which is MJD 58131. All epochs are transformed to the rest-frame ones.}} &\colhead{$T_{\rm ph}$}&\colhead{$R_{\rm ph}$}&\colhead{$T_{\rm d}$}&\colhead{$M_{\rm d}$}&\colhead{$\chi^{\rm 2}$/dof}&\colhead{$T_{\rm ph}$}&\colhead{$R_{\rm ph}$}&\colhead{$T_{\rm d}$} & \colhead{$M_{\rm d}$}  & \colhead{$\chi^{\rm 2}$/dof} \\
		(days) & (K) & (10$^{15}$ cm) &  (10$^{3}$\,K) & (10$\rm ^{-3}\ M_{\odot}$) &    & (K) & (10$^{15}$ cm) & (10$^{3}$\,K) & (10$\rm^{-3}\ M_{\odot}$) & \\\hline
		120.2d	&$	5577.90 	^{+	111.00 	}_{-	105.84 	}$&$	0.77 	^{+	0.03 	}_{-	0.03 	}$&$	730.97 	^{+	114.12 	}_{-	61.60 	}$&$	19.48 	^{+	43.06 	}_{-	15.93 	}$&	5.89 	&$	5578.90 	^{+	 113.03 	}_{-	106.33 	}$&$	0.78 	^{+	0.04 	}_{-	0.03 	}$&$	778.38 	^{+	111.61 	}_{-	59.92 	}$&$	24.97 	^{+	42.12 	}_{-	19.36 	}$&	5.95 	\\\hline
		137.83d	&$	5662.40 	^{+	132.20 	}_{-	125.73 	}$&$	0.62 	^{+	0.03 	}_{-	0.03 	}$&$	782.24 	^{+	209.56 	}_{-	114.32 	}$&$	5.03 	^{+	33.11 	}_{-	4.64 	}$&	6.78 	&$	5663.89 	^{+	 136.42 	}_{-	127.31 	}$&$	0.62 	^{+	0.03 	}_{-	0.03 	}$&$	829.07 	^{+	220.23 	}_{-	115.19 	}$&$	6.99 	^{+	36.10 	}_{-	6.37 	}$&	6.82 	\\\hline
		161.92d	&$	6033.30 	^{+	294.80 	}_{-	256.57 	}$&$	0.41 	^{+	0.05 	}_{-	0.04 	}$&$	846.32 	^{+	249.12 	}_{-	157.24 	}$&$	2.09 	^{+	22.43 	}_{-	1.92 	}$&	1.09 	&$	6040.78 	^{+	 304.24 	}_{-	254.25 	}$&$	0.41	^{+	0.05 	}_{-	0.04 	}$&$	886.62 	^{+	257.65 	}_{-	156.72 	}$&$	3.55 	^{+	28.87 	}_{-	3.22 	}$&	1.10 	\\\hline
	\end{tabular}
\end{table*}

\clearpage

\clearpage

\clearpage

\clearpage

\clearpage

\clearpage

\clearpage

\clearpage

\clearpage

\appendix
\setcounter{table}{0}
\setcounter{figure}{0}
\setcounter{equation}{0}
\renewcommand{\thetable}{A\arabic{table}}
\renewcommand{\thefigure}{A\arabic{figure}}
\renewcommand\theequation{A.\arabic{equation}}

\begin{figure}[tbph]
\begin{center}
\includegraphics[width=0.9\textwidth,angle=0]{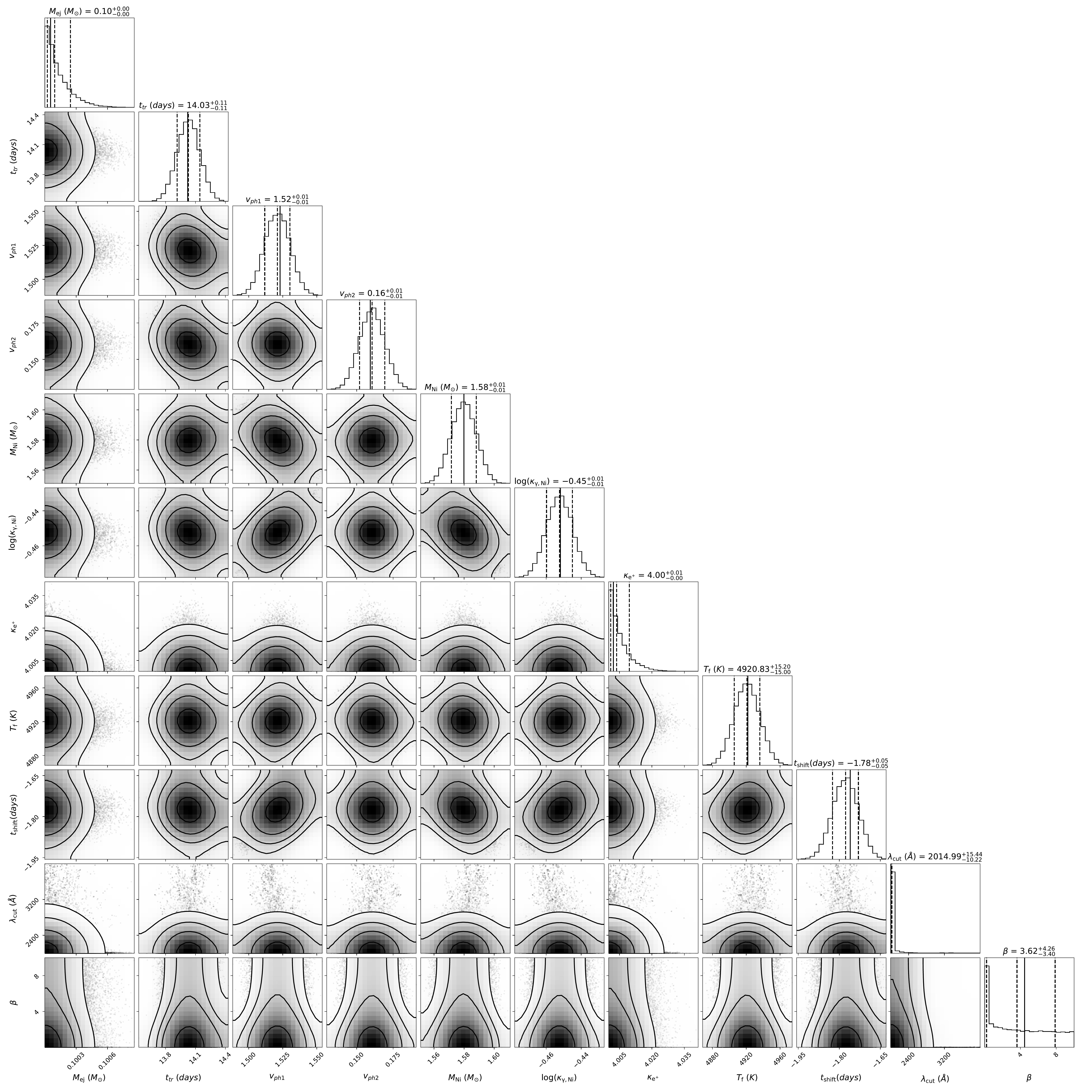}
\end{center}
\caption{The corner plot of the \Ni model. The solid vertical lines represent the best-fitting parameters, while the dashed vertical lines
represent the medians and the 1-$\sigma$ bounds of the parameters.}
\label{fig:corner_Ni}
\end{figure}

\begin{figure}[tbph]
\begin{center}
\includegraphics[width=0.9\textwidth,angle=0]{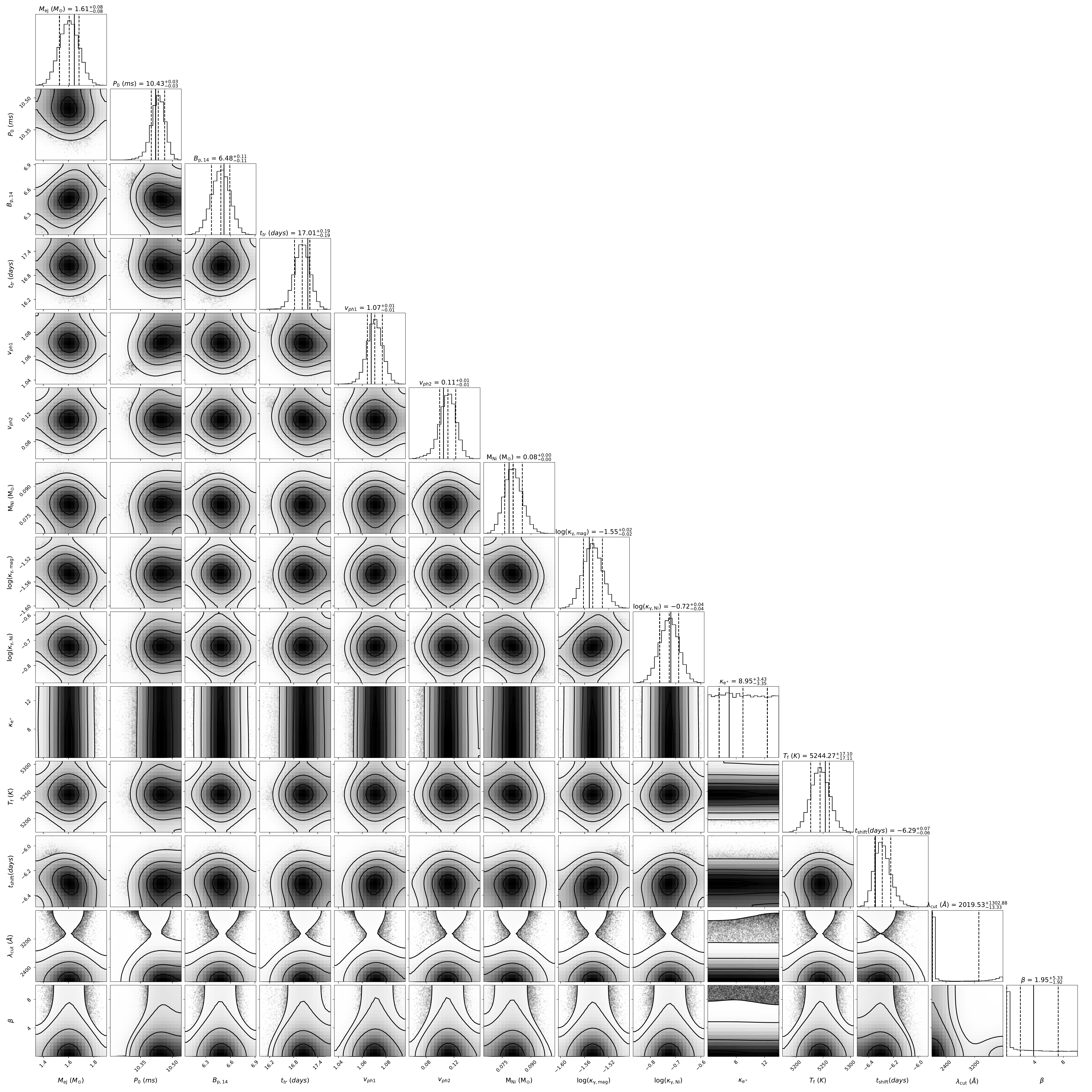}
\end{center}
\caption{The corner plot of the magnetar plus \Ni model. The solid vertical lines represent the best-fitting parameters, while the dashed vertical lines
represent the medians and the 1-$\sigma$ bounds of the parameters.}
\label{fig:corner_Mag+Ni}
\end{figure}

\begin{figure}[tbph]
	\begin{center}
		\includegraphics[width=0.9\textwidth,angle=0]{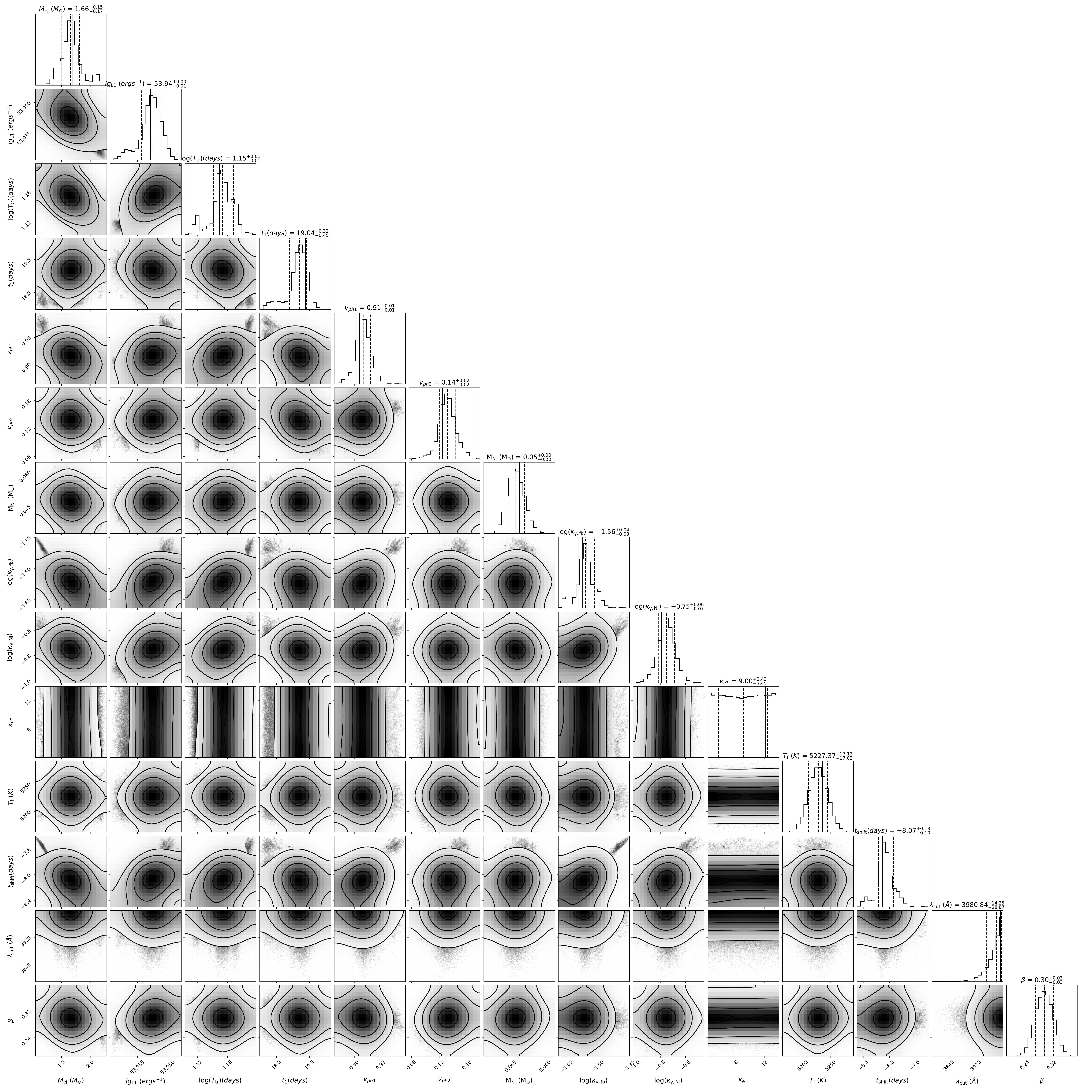}
	\end{center}
	\caption{The corner plot of the fallback plus \Ni model. The solid vertical lines represent the best-fitting parameters, while the dashed vertical lines
		represent the medians and the 1-$\sigma$ bounds of the parameters.}
	\label{fig:corner_fallback+ni}
\end{figure}

Figures \ref{fig:corner_Ni}, \ref{fig:corner_Mag+Ni}, and \ref{fig:corner_fallback+ni}
present the corner plots of the \Ni model, the magnetar plus \Ni model, and the fallback plus \Ni model, respectively.

\clearpage

\clearpage

\end{document}